\documentclass[prd,amsfonts,aps,nofootinbib,12pt]{revtex4-1}
\pdfoutput=1
\usepackage{amsmath,amssymb}
\usepackage{graphicx}
\usepackage[utf8]{inputenc}
\usepackage{cancel}
\usepackage{slashed}
\usepackage[colorlinks=true]{hyperref}
\usepackage{color}

\newcommand{\EIAaddress}{Grupo F\'isica Te\'orica y Aplicada, Universidad EIA, A.A. 7516, Medell\'in, Colombia,}
\newcommand{\UdeAaddress}{Instituto de F\'isica, Universidad de Antioquia, Calle 70 No. 52-21, Medell\'in, Colombia.}
\begin{document}
%%%%%%%%%%%%%%%%%%%%%%%%%%%%%%%%%%%%%%%%%%%%%%%%%
\title{Phenomenology of doublet-triplet fermionic dark matter in nonstandard cosmology and multicomponent dark sectors}

\author{Amalia Betancur${}^{1,2}$}\email{amalia.betancur@eia.edu.co}
\author{Óscar Zapata${}^2$}\email{oalberto.zapata@udea.edu.co}

\affiliation{$^1$~\EIAaddress}
\affiliation{$^2$~\UdeAaddress}
\date{\today}

%%%%%%%%%%% ABSTRACT %%%%%%%%%%%%%%%%%%%%%%%%%%%%%%%%%%%%%%
\begin{abstract}
  We consider the doublet-triplet fermionic model  in the scheme of the custodial limit when the dark matter (DM) candidate is pure doublet and lies at the electroweak scale. %, which is a simplified extension of the SM with an additional vectorlike $SU(2)_L$ doublet and a Majorana $SU(2)_L$ triplet. 
  This scheme, despite being an appealing and promising DM model, is severely constrained by the interplay between the DM relic density constraint and the LHC measurement of the Higgs diphoton decay rate. 
  In this work, the DM relic density is considered to arise from either a nonstandard cosmology scenario or as a part of a larger sector encompassing other DM particles, this in order to saturate the observed relic abundance.
  For these scenarios we investigate the impact of the new sector in different collider observables, and study constraints coming from direct detection and indirect detection of gamma-rays both in the diffuse and linelike spectrum.
  As a result, we find that in the nonstandard cosmology scenario most experiments impose, up to a certain point, restrictions, though large portions of the parameter space are still viable. For the multicomponent case, only direct detection imposes constraints.

\end{abstract} 

\maketitle

\section{Introduction} 
As early as 1930s, astronomical observations hinted the existence of an unknown form of matter. In the last decades the evidence for the existence of this type of matter, dubbed dark matter (DM) \cite{Zwicky:1933gu}, has become abundant and overwhelming \cite{Tanabashi:2018oca}.
We now know that it makes up about 80$\%$ of the universe matter-content \cite{Aghanim:2018eyx} pointing out to the existence of one or several new particles that are not part of the Standard Model (SM) of particle physics.
Among the DM canditates \cite{Feng:2010gw}, weakly interacting massive particles (WIMPs) \cite{Steigman:1984ac} are some the most well-motivated candidates since the thermal annihilation cross section needed to account for the observed DM relic density is obtained for DM particles with electroweak interactions and masses.
That is, the WIMPs lie naturally at what is expected to be the scale of physics beyond the Standard Model (BSM).
Moreover, their abundance \cite{Steigman:2012nb} is governed by the generic mechanism of chemical frezee-out which has also played a role in the abundance of light elements as well as the cosmic microwave background radiation \cite{Kolb:1990vq}, both of which are in stark agreement with current observations.

It is worth noting, however, that the WIMP paradigm is not free of challenges, both at theoretical and experimental levels \cite{Baer:2014eja}.
For instance, it is not always a given that a WIMP candidate will automatically account the total DM relic abundance, which, in some cases, implies the need for some degree of fine-tuning in order for the models to still be viable\footnote{In particular, one of the most well studied WIMPs, the neutralino, tends to yield too much relic abundance if it is  bino-like while its relic abundance tends to be suppressed (as long as its mass is below 2.4 TeV) if it is mainly wino \cite{Jungman:1995df,Bertone:2004pz}.}.
Additionally, despite large efforts to find evidence of WIMPs through production at colliders, elastic scattering with heavy nuclei while passing through the Earth or observation of the self-annihilation byproducts in regions with high DM density, no concluding evidence has been reported.
The null results have lead to more and more constraints on the parameter space of popular WIMP models \cite{Bertone:2010at,Baer:2014eja,Escudero:2016gzx,Arcadi:2017kky}.

Among the different approaches to overcome the challenges on WIMP models, we recall those that depart completely or partially from  standard cosmology scenario  \cite{Kamionkowski:1990ni,Giudice:2000ex,Gelmini:2006pw,Acharya:2009zt} or consider WIMPs to make up only a fraction of the total DM of the Universe \cite{Zurek:2008qg,Profumo:2009tb}. 
In standard cosmology, the WIMP relic density is calculated assuming a set conditions during the time before Big Bang Nucleosynthesis (BBN), but there are no data or indirect evidence supporting such assumptions.
For instance, the reheating temperature is an unknown quantity which could be as low as $\sim1$ MeV, the temperature at which nucleosynthesis begins. If this temperature is small, it may have a profound implication on the DM relic density (be suppressed or enhanced compared to the standard scenario) \cite{Giudice:2000ex,Gelmini:2006pw}.
More recently, several works have addressed the problem of DM abundance under nonstandard cosmologies \cite{Aparicio:2016qqb,DEramo:2017gpl,Hamdan:2017psw,Visinelli:2017qga,DEramo:2017ecx,Bernal:2018ins,Bernal:2018kcw,Hardy:2018bph,Drees:2017iod,Arbey:2018uho}.
For instance, in Refs.~\cite{Bernal:2018kcw,Hardy:2018bph} the case for scalar DM was addressed, while in Ref.~\cite{Drees:2017iod} a generic calculation for the DM abundance with a late decaying scalar was considered, and in Ref.~\cite{Arbey:2018uho} the relic abundance is considered as a probe of the conditions of the Universe pre-BBN. 
Interestingly enough, such deviations from the standard cosmology do not affect the prospects regarding DM detection and BBN. 
On the other hand, although most proposals contain one single WIMP, there is no reason to consider that the DM of the Universe is composed by just one type of DM particles. 
The total DM relic abundance could be a result of the presence of several DM particles, a scenario referred to as multicomponent DM.
In such scenarios, the DM of the Universe is set by the WIMP and other DM candidate, such as QCD axions \cite{Baer:2011hx,Bae:2013hma,Dasgupta:2013cwa,Alves:2016bib,Ma:2017zyb} or even another WIMP ~\cite{Zurek:2008qg,Profumo:2009tb,Esch:2014jpa,Arcadi:2016kmk,Bernal:2018aon}.
In light of this, it makes sense to study models where the relic abundance is not imposed as a constraint, either because in nonstandard cosmology it is possible to fulfil this requirement when the right combination of parameters is achieved, or because is achieved by the interplay of two separate dark sectors. 

For the aforementioned reasons, we consider the doublet-triplet fermionic model (DTF) \cite{Dedes:2014hga,Abe:2014gua,Freitas:2015hsa,Lopez-Honorez:2017ora}, which is one of the ultraviolet realizations of the fermionic Higgs portal \cite{Patt:2006fw}, and is part of the minimal setup expected when the SM is extended by new physics which is, to some extent, related to lepton and baryon number conservation~\cite{Arbelaez:2015ila,Arkani-Hamed:2015vfh}. 
In the DTF, the SM is enlarged with two colorless fermions, a vectorlike $SU(2)_L$ doublet and a majorana $SU(2)_L$ triplet, both being odd an exact $Z_2$ symmetry\footnote{This symmetry can be recognized as remnant symmetry at the end of the symmetry breaking chain of the SO(10) group to the SM \cite{Arbelaez:2015ila}.} which is imposed in order to render stable the lightest $Z_2$-odd particle\footnote{This particle setup has been also considered in studies associated to strengthening the electroweak phase transition \cite{Carena:2004ha}, presicion test in future colliders \cite{Arkani-Hamed:2015vfh,Bertuzzo:2017wam,Xiang:2017yfs}, electroweak precision tests \cite{Cai:2016sjz}, generation of neutrino masses \cite{Betancur:2017dhy}, fake split-supersymmetry \cite{Benakli:2013msa}, and UV completion of doublet fermion DM \cite{Dedes:2016odh}.}.
The DM candidate turns out to be a mixture, generated by the interaction with the Higgs boson, between the neutral component of the triplet  and the neutral components of the doublet vectorlike fermion. 
The viable dark matter regions comprise a DM mass around the electroweak scale and above 1 TeV \cite{Dedes:2014hga,Abe:2014gua,Freitas:2015hsa,Betancur:2017dhy}.
The electroweak DM region arises in the scheme of the custodial limit (the new Yukawa couplings are equal) when the DM candidate is pure doublet.  However, the contribution of new the charged fermions to $h \rightarrow \gamma  \gamma$  generates a considerable suppression on the Higgs diphoton decay making such a scheme severely constrained by the interplay between the DM relic density constraint and the LHC measurement of the Higgs diphoton decay rate \cite{Dedes:2014hga,Abe:2014gua,Freitas:2015hsa,Betancur:2017dhy}. 

In this work we will study the custodial limit of the DTF within either a nonstandard cosmology scenario, {\it i.e.}, we depart from the standard relic density calculation, or a multicomponent DM  setup but assuming that the WIMP relic density is obtained through the thermal freeze out. 
We will establish the current constraints on the DFT coming from collider searches and Higgs diphoton decay, without taking care of the constraint on the DM relic density.
Then, we will go on to determine the restrictions resulting from direct detection (DD) experiments and indirect detection (ID) with gamma rays, both in the framework of nonstandard cosmology and the DM candidate as part of a multicomponent system.

The rest of the paper is organized as follows. In Sec. \ref{themodel} we present the doublet-triplet fermion model with its mass spectra and discuss the allowed interactions. In Sect. \ref{collider} we present the model restrictions arising  from electroweak  production at colliders of charginos and neutralinos and from precision measurements of the Higgs diphoton decay rate. In Section \ref{NONSC} we study the constraints arising from DD and ID with gamma rays for the case of a nonstandard cosmology, whereas in Section \ref{multiDM}, the same analysis is made for the case where the DM candidate is part of a multicomponent system. Finally, we conclude in Sec.~\ref{sec:conc}.

\section{The Model}\label{themodel}
Doublet-triplet fermion DM consists on extending the SM with an $SU(2)_L$ vectorlike doublet with $Y=-1$ and a Majorana $SU(2)_L$ triplet, both odd under an exact $Z_2$ symmetry. 
Expressing the new $SU(2)_L$ multiplets as
\begin{align}\label{eq:fermioncontent}
\psi=\left( \begin{array}{ccc}
\psi^0  \\
\psi^- \end{array} \right),\hspace{1cm} 
\Sigma_L=\left( \begin{array}{ccc}
 \Sigma_L^0/\sqrt{2} &  \Sigma_L^+\\
 \Sigma_L^{-} & -\Sigma_L^0/\sqrt{2} \end{array} \right),
\end{align}
the most general renormalizable Lagrangian, invariant under the $SU(2)_L \times U(1)_Y \times Z_2$ symmetry can be written as 
\begin{align}
\mathcal{L}&=\mathcal{L}_{\rm{SM}}+\mathcal{L}_{\rm{F}}+\mathcal{L}_{\rm{I}}-\mathcal{V}_{\rm{SM}}.
\end{align}

Here $\mathcal{L}_{\rm{SM}}$ is the SM Lagrangian and $\mathcal{V}_{\rm{SM}}$ is the scalar potential of the Higgs doublet $H=(0, (h+v)/\sqrt{2})^{\text{T}}$,  with $h$ being the Higgs boson and $v=246$ GeV.
$\mathcal{L}_{\rm{F}}$ refers to the kinetic and mass terms of the new fermions,
\begin{align}
\mathcal{L}_F&=\bar{\psi} i \cancel{D}\psi-M_\psi\bar{\psi}\psi+{\rm Tr}[\bar{\Sigma}_L i\cancel{D} \Sigma_L]-\frac{1}{2}{\rm Tr}(\bar{\Sigma}_L^cM_\Sigma\Sigma_L+\mbox{h.c.}),
\end{align}
and $\mathcal{L}_{\rm{I}}$  contains the Yukawa interactions of the new fermions with the Higgs doublet,
\begin{align}\label{eq:LI}
\mathcal{L}_{\rm{I}}= -y_1 H^\dagger\bar{\Sigma}_L^c\epsilon \psi_R^c   +  y_2 \bar{\psi}_L^c \epsilon \Sigma_L H  + {\rm h.c.},
\end{align}
where $y_1$ and $y_2$ are new Yukawa couplings that generate the mixing between the new fermion fields.
Note that the $Z_2$ symmetry not only guarantees the stability of the lightest $Z_2$-odd particle (the DM particle) but also avoids Higgs-mediated flavor changing neutral currents at tree level through the mixing terms $\bar{\psi}He_R$ and $\overline{\Sigma}^c_L\tilde{H}^\dagger L$. Therefore, lepton flavor violating processes such as $\mu\to e\gamma$ are forbidden.%  In other words, the lepton flavor number remains conserved in our model.

The particle spectrum contains three new Majorana mass eigenstates $\chi_{\alpha}^0$ ($\alpha=a,b,c$) and two new charged fermions $\chi_{a,b}^{\pm}$\,.
The mass matrix for the neutral fermions is \cite{Dedes:2014hga,Betancur:2017dhy} (in the basis $\Xi^0=(\Sigma_L^0, \psi^0_L, \psi^{0c}_R)^T$), 
\begin{align}
\label{eq:MchiN}
  \mathbf{M}_{\Xi^0}=\begin{pmatrix}
 M_\Sigma                 &\frac{1}{\sqrt{2}}yv\cos\beta& \frac{1}{\sqrt{2}}yv\sin\beta\\
\frac{1}{\sqrt{2}}yv\cos\beta &  0                  & M_\psi\\
\frac{1}{\sqrt{2}}yv\sin\beta&  M_\psi                &  0
\end{pmatrix},
\end{align}
with $y=\sqrt{(y_1^2 + y_2^2)/2}$ and $\tan\beta=y_2/y_1$.
On the other hand, the charged fermions mass matrix reads
\begin{align}
\label{eq:MchiC}
\hspace{1cm}
  \mathbf{M}_{\Xi^\pm}=\begin{pmatrix}
 M_\Sigma         &   yv\cos\beta \\
 yv\sin\beta & M_\psi \\
\end{pmatrix},
\end{align}
which is diagonalized by a rotation of the gauge eigenstates into the physical states defined via
\begin{align}
\left( \begin{array}{cc}
\Sigma^+ \\
\psi^+
\end{array} \right)
= %
\left( \begin{array}{cc}
\cos\theta & -\sin\theta \\
\sin\theta & \cos\theta
\end{array} \right)
\left( \begin{array}{cc}
\chi_a^+\\
\chi_b^+
\end{array} \right), \hspace{2cm}
\sin(2 \theta)= \frac{\sqrt{2} \ y \ v}{m_{\chi_a^+}-m_{\chi_b^+}^2}.
\end{align}
Either one of the mass eigenstates $\chi_a^{\pm}$ and $\chi_b^{\pm}$ could be the lightest charged $Z_2$-odd fermion.
From now on, we will assume that $\chi_1^{\pm}$ is the lightest between $\chi_a^{\pm}$ and $\chi_b^{\pm}$, while $\chi_2^{\pm}$ is the heaviest.
Thus, for the masses $m_{\chi_1^{\pm}}$ and $m_{\chi_2^{\pm}}$ a mass ordering is implied.
On the other hand, these mass matrices evoke the neutralino and chargino mass matrices (in the wino-higgsino limit) in the minimal supersymmetric standard model \cite{Martin:1997ns}, which case is realized when $y=g/\sqrt{2}$ and has been exploited in studies such as \cite{Carena:2004ha,Arkani-Hamed:2015vfh,Benakli:2013msa}. 

\subsection{The custodial limit}
An interesting scheme arises when $\tan\beta=1$ (the custodial limit) since several consequences arise. 
First, the interaction Lagrangian $\mathcal{L}_{\rm{I}}$ becomes invariant under an $SU(2)_R$ symmetry which protects the $T$ and $U$ parameters \cite{Cai:2016sjz, Dedes:2014hga}, thus making the model free of the constraints coming from electroweak precision tests.
Second, all diagonal tree-level couplings of $\chi_\alpha^0$ to the $Z$ boson are zero.   
And third, the neutral mass matrix may be written as\footnote{ This can be done via a similarity transformation  $\mathbf{M}'_{\Xi^0}=O^\dagger\mathbf{M}_{\Xi^0}O$, with
    \begin{align}
      O=\begin{pmatrix}
        1  & 0 & 0\\
        0 & \frac{1}{\sqrt{2}} & -\frac{1}{\sqrt{2}} \\
        0 &\frac{1}{\sqrt{2}} & \frac{1}{\sqrt{2}} \\
      \end{pmatrix}.
    \end{align}.}
  \begin{align}\label{eq:Mchi2}
    \mathbf{M}'_{\Xi^0}=\begin{pmatrix}
      M_\Sigma                 &yv& 0\\
      yv &  M_\psi                  & 0\\
      0 &  0                &  -M_\psi  \\
    \end{pmatrix},
  \end{align}
  which shows that there is a decoupled eigenvalue $M_\psi$ (the mass is not obtained from the electroweak symmetry breaking), meaning that one of the mass eigenstates (which will be labelled as $\chi_1^0$ with mass $m_{\chi_1^0}=-M_\psi$) is an equal mixture of the doublets with no triplet component.
Moreover, the coupling of $\chi_1^0$ to the Higgs $(g_{h\chi_1^0\chi_1^0 })$ is also zero at tree level.

It follows that, for the custodial limit scheme with $\chi_1^0$ as the DM candidate ($\chi_1^0$ is the lightest between $\chi_a^0, \chi_b^0$ and $\chi_c^0$)\footnote{In this case we have that the heavier neutral fermions are mass-degenerate with the charged fermions ($m_{\chi_2^0}= m_{\chi_1^\pm}$ and $m_{\chi_3^0}= m_{\chi_2^\pm}$), so $|m_{\chi_1^0}| < |m_{\chi_2^0}| < |m_{\chi_3^0}|$.}, the above features lead to the following important implications for the phenomenology of the model.   
\begin{itemize} 
\item   DM ($\chi_1^0$) annihilates into weak gauge bosons through $t$- and $u$- channels via exchange of heavier $Z_2$ fermions, and so, the DM annihilation cross section is suppressed.
  Furthermore, for large Yukawa couplings\footnote{This is the reason for which such a scheme can not be obtained in the MSSM.} $(0.5\lesssim y\lesssim3)$, the splitting between the DM candidate and the heavier new fermions is large which further suppresses the annihilation cross section, thus allowing the DM candidate to saturate the relic abundance for masses as low as the electroweak scale ($80\lesssim m_{\chi_1^0}\lesssim200$ GeV).
\item Within the region where the correct DM relic density is obtained, there are two different allowed triplet mass regions for any value of the pair $y$-$m_{\chi_1^0}$, one where $M_{\Sigma}$ is always negative ($M_{\Sigma}<-m_{\chi_1^0}$) and one where it can be either positive or negative but larger than $-m_{\chi_1^0}$. 
\item Since the interactions between $\chi_1^0$ and both $h$ and $Z$ are zero a tree-level, there are not contributions to direct detection at tree-level, so the leading contributions appear at one-loop level (more on this in Sec. \ref{DD}) for both spin-independent (SI) and spin-dependent cross sections. These blind spots for the model has been studied in works such as \cite{Dedes:2014hga,Freitas:2015hsa,Betancur:2017dhy}
\item Though the custudial limit scheme presents interesting possibilities, it is also severely constrained by the Higgs diphoton decay measurement.
  The origin of this is due to the presence of new charged fermions that couple to the Higgs boson and create an effective $h\gamma\gamma$ coupling through a triangular diagram similar to the top quark contribution (though with a larger electric charge).
  The decay ratio with respect to the SM rate can be parametrized as
  \begin{align}
    \label{eq:higgsgammagamma}
    &R_{\gamma\gamma}= \left|1 + \frac{1}{A_{SM}}\left[ \frac{y^2v^2}{m_{\chi_2^\pm}-m_{\chi_1^\pm}}\left(\frac{A_F(\tau_{\chi_2^\pm})}{m_{\chi_{2}^{\pm}}}-\frac{A_F(\tau_{\chi_1^\pm})}{m_{\chi_{1}^{\pm}}}\right)\right]  \right|^2,
  \end{align}
  where $A_{{\rm SM}}=-6.5$ includes the contribution from all charged SM particles such as gauge bosons and fermions, and the loop factor is $A_F(\tau)= 2 \tau ^{-2}[\tau + (\tau-1)\arcsin^2{\sqrt{\tau}}]$ for $\tau\leq1$ where $\tau_X=m^2_{h}/(4 m^2_{X})$.
  As can be seen from Eq.~(\ref{eq:higgsgammagamma}), the new fermion contribution is always positive for real Yukawa couplings (which is the case at hand) and because it is opposite in sign to SM contribution, it may suppress its value causing large deviations from the results published by the ATLAS~\cite{Aaboud:2018xdt} and CMS~\cite{Sirunyan:2018ouh} collaborations.
  
  Indeed, that is what occurs when large values of $y$ as well as large splitting between the $Z_2$ odd heavier fermions are required, the same conditions that lead to the correct DM relic abundance.
  As a result, the new contribution conspirates to create large discrepancies from the expected value, up to the point of only allowing the DM mass to lie between the narrow range of $70 < m_{\chi_1^0}< 80$ (GeV).
\end{itemize}
All in all, despite the custodial limit of the DTF model being an appealing and promising scenario (thus being an excellent exponent of the WIMP paradigm), it is severely constrained by the interplay between the DM relic density constraint and the LHC measurement of the Higgs diphoton decay rate. 

Nonetheless, if the DM abundance is generated within a nonstandard cosmology or is part of multicomponent dark sector, the requirement of having large values for $y$ as well as a large mass splitting between the charged fermions would be substantially weaken.
This in turn would result in a larger portion of the parameter space that is still allowed\footnote{Other way of recovering  part of the parameter space is via introduction of additional charged scalars \cite{Betancur:2017dhy}.}.

In the following sections, we will explore the DTF in the case of $\tan \beta=1$ with $\chi_1^0$ as the DM candidate, our aim is to find current as well as upcoming restrictions due to collider, direct detection and indirect detection experiments.
We choose the set of free parameters of the model to be $M_\psi, M_\Sigma$ and $y$, and due to the freedom to make field redefinitions we assume $M_\psi,y>0$ and $M_\Sigma$ to be real \cite{Dedes:2014hga}, implying CP conservation in the $Z_2$-odd sector and that three intrinsic CP phases of the $Z_2$-odd fields (including $\chi_1^0$) are fixed. 

\section{Collider Bounds}\label{collider}
The Large Hadron Collider (LHC) is currently running at an outstanding 13 TeV energy and has collected more that 100 fb$^{-1}$ of data. One of its current goals is to probe BSM models either by direct production of new particles or by measuring possible deviations from the SM. In that regard, the parameter space of the DTF model may be constrained by the ATLAS and CMS experiments. 
\subsection{Higgs diphoton decay}\label{sec:diphoton}
\begin{figure}[t!]
  \begin{center}
    % \label{fig:Rgg-md-y-MT}
    \includegraphics[scale=0.35]{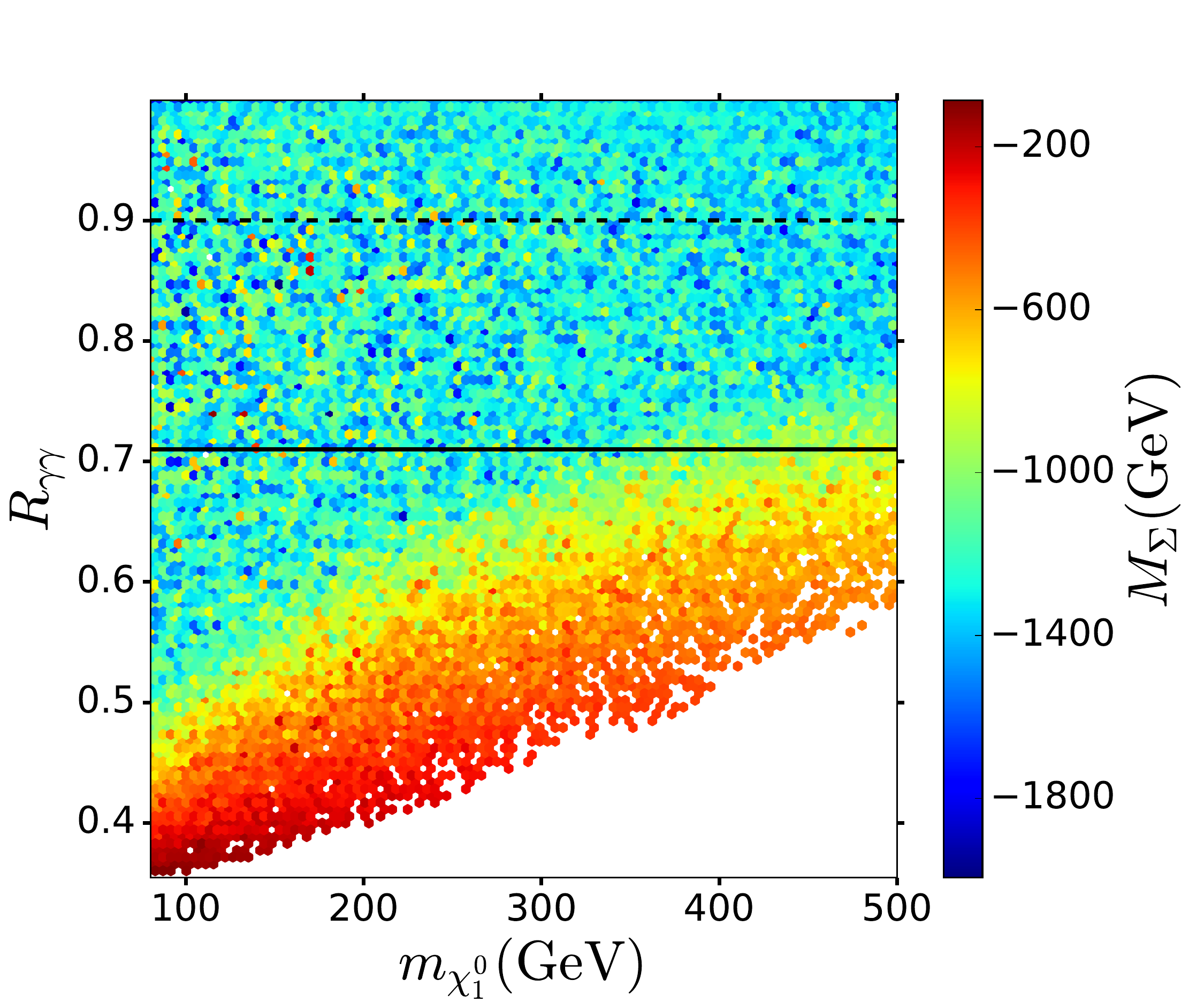}
    \includegraphics[scale=0.35]{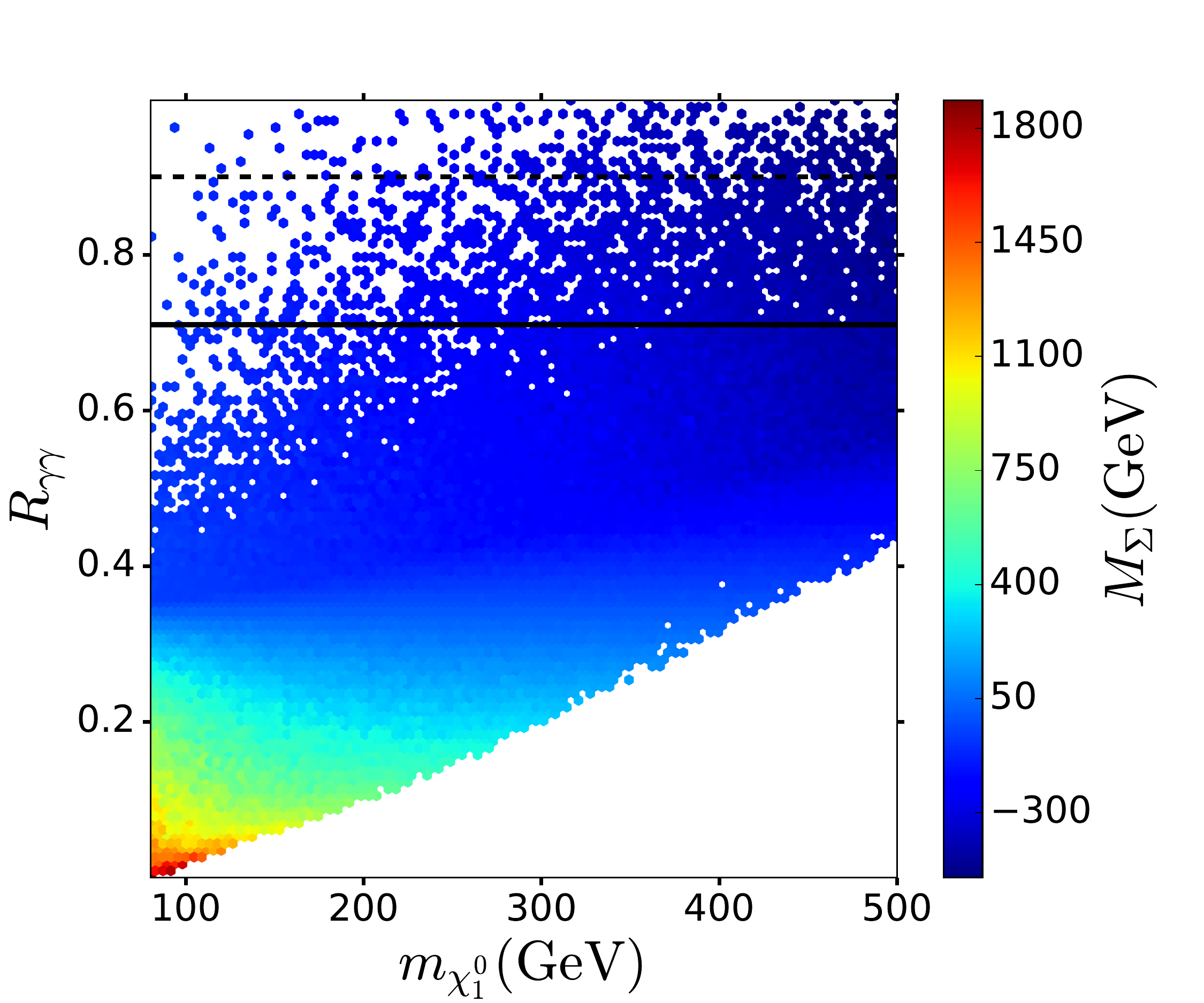}\\
    \includegraphics[scale=0.35]{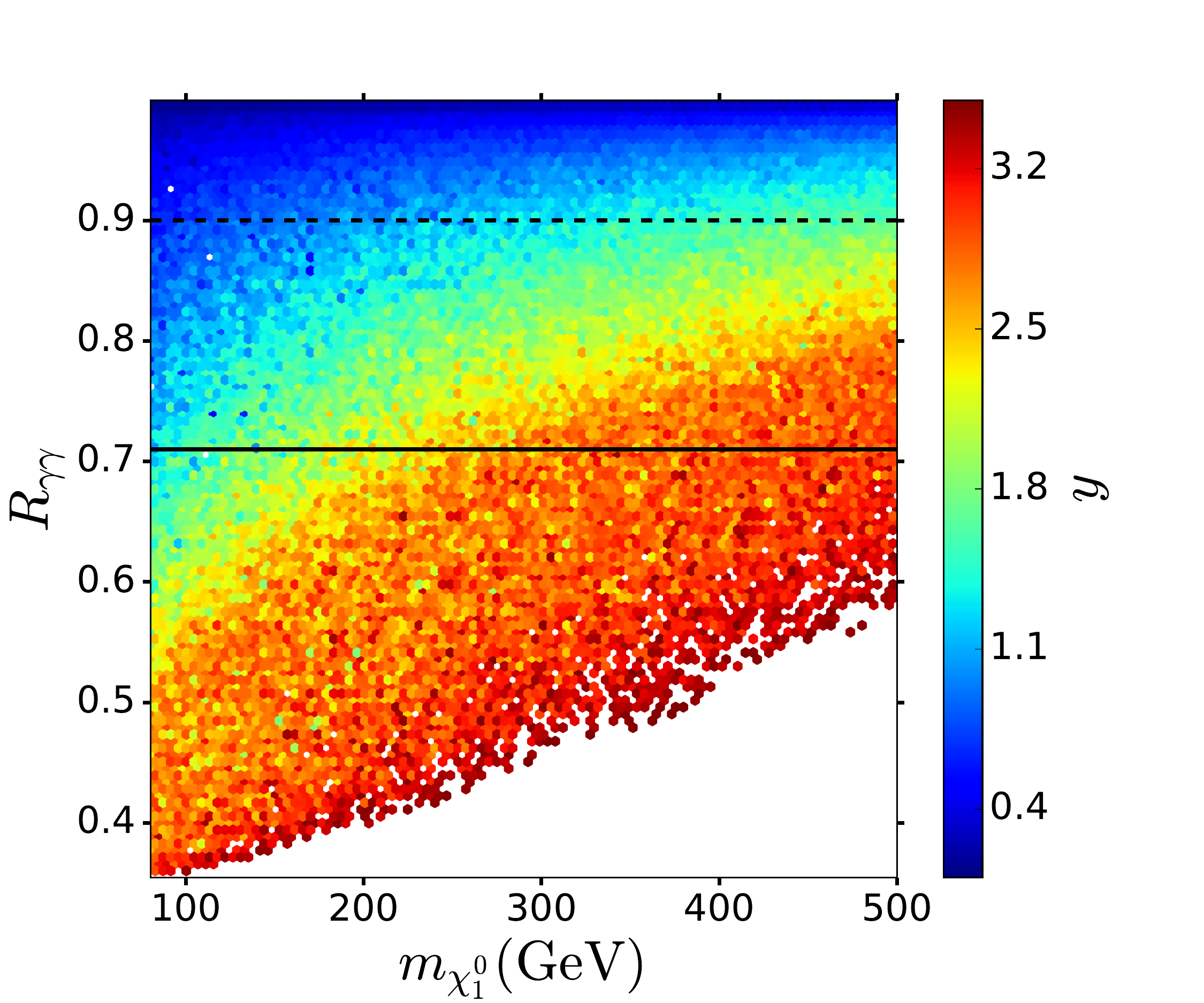}
    \includegraphics[scale=0.35]{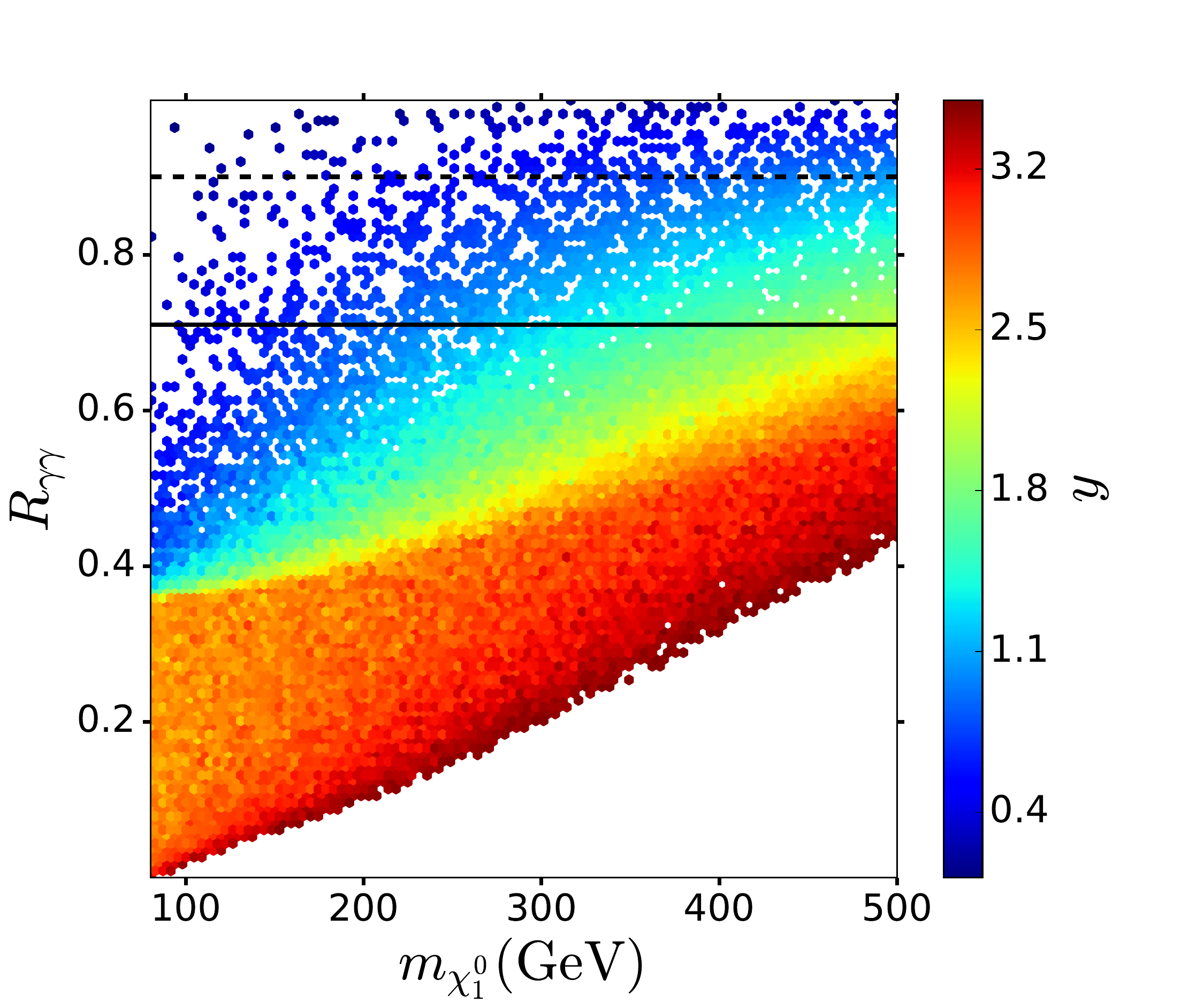}
    \caption{Scan on the parameter space of the model against $R_{\gamma \gamma}$ for the region where $M_{\Sigma}<-m_{\chi^0_1}$ (left panels) and the region where $M_{\Sigma}>-m_{\chi^0_1}$  (right panels). The solid and dashed horizontal lines represent the lowest bound at a 2$\sigma$ deviation from the central value reported by the ATLAS and CMS collaboration respectively. }
    \label{fig:Rgg-1}
  \end{center}
\end{figure}

As explained before, the DTF may generate large deviations from the current measurements of the Higgs diphoton decay rate.
The ATLAS and CMS collaborations have presented results for the decay at $\sqrt{s}=13$ TeV with $\sim 36\, \mathrm{fb^{-1}}$ of data in Refs.~\cite{Aaboud:2018xdt} and CMS \cite{Sirunyan:2018ouh}, respectively.
We use these results to find the regions of the parameter space that are in agreement with the Higgs diphoton decay rate. For this we consider models that deviate at most $2\sigma$ from the central value reported by each collaboration.

In order to obtain constraints from $R_{\gamma \gamma}$ without imposing the DM relic density constraint,  we performed a scan of the free parameters of the model as follows:
\begin{align}\label{eq:scan}
  0.1 < &\; y<3.5,\nonumber\\
  -2000 \ \mathrm{GeV} < &\; m_{\Sigma}<2000\,\mathrm{GeV},\nonumber\\
  75.0 \ \mathrm{GeV} < &\; m_{\psi}<500 \ \mathrm{GeV}. 
\end{align}
Additionally, we only considered models where the lightest charged fermion is heavier than 100 GeV in order to satisfy LEP limits \cite{Achard:2001qw}.

The results are presented in Fig.~\ref{fig:Rgg-1}  where the parameter space has been divided into two regions, $M_{\Sigma}<-m_{\chi^0_1}$ (left panels) and $M_{\Sigma}>-m_{\chi^0_1}$ (right panels). We will do this throughout the paper because the phenomenology in these two regions tend to yield different results.
The scan shows that, considering the ATLAS results, for the region where $M_{\Sigma}<-m_{\chi^0_1}$ there are no restrictions on the Yukawa coupling $y$ or on $M_{\Sigma}$ whereas in the region $M_{\Sigma}>-m_{\chi^0_1}$ the decay rate forbids $y$ values larger than 2.25 and $M_{\Sigma}\lesssim -60$ GeV. 
On the other hand, CMS results yield the severe constraint $y<2.0$ for $M_{\Sigma}<-m_{\chi^0_1}$ and $y<1.0$ for $M_{\Sigma}>-m_{\chi^0_1}$, whereas $M_{\Sigma}$ must be less than $\sim-92$ GeV. Hence, positive triplet masses are no longer consistent with $R_{\gamma \gamma}$ results.

\begin{figure}[t!]
\begin{center}
\includegraphics[scale=0.35]{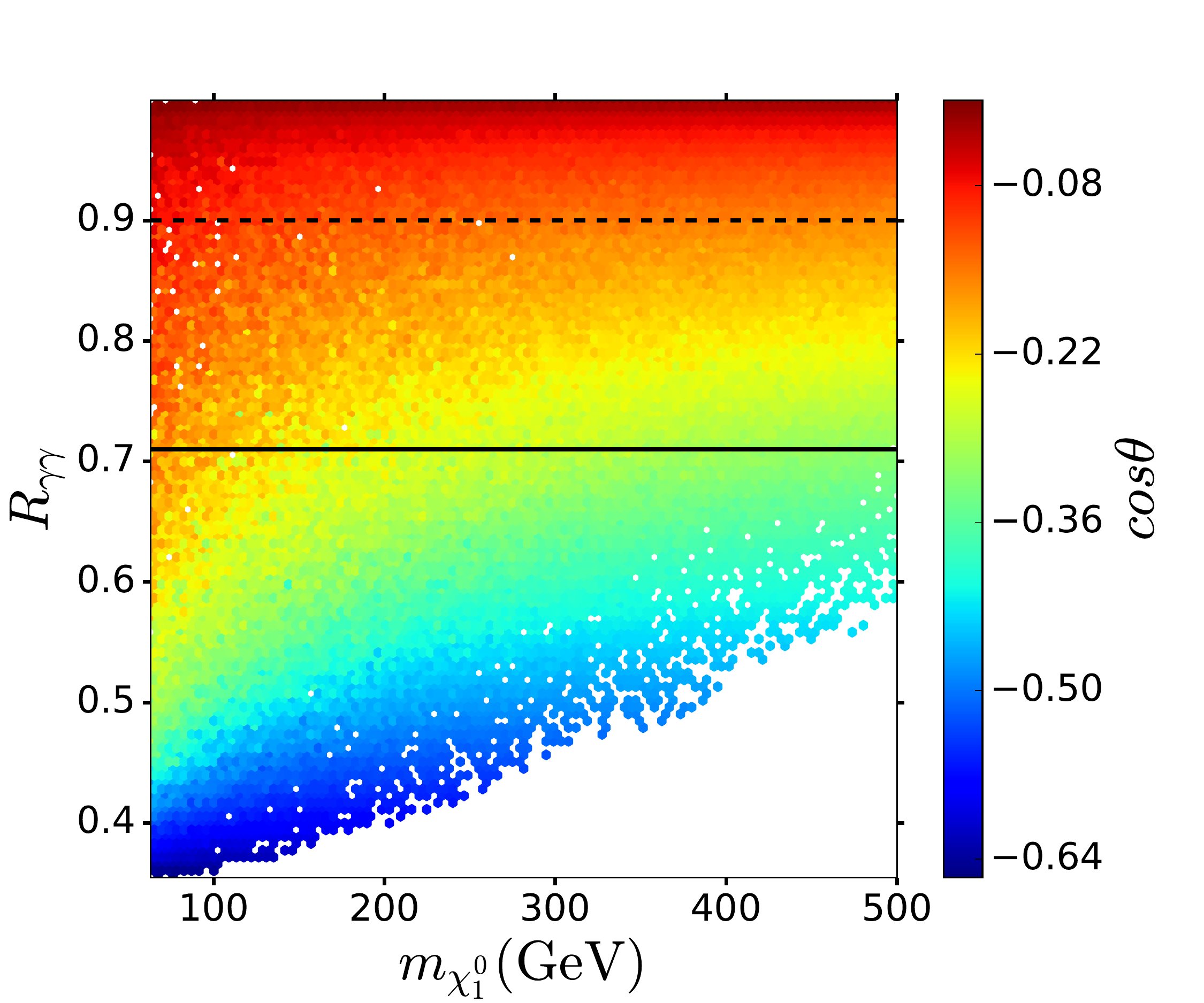}
\includegraphics[scale=0.35]{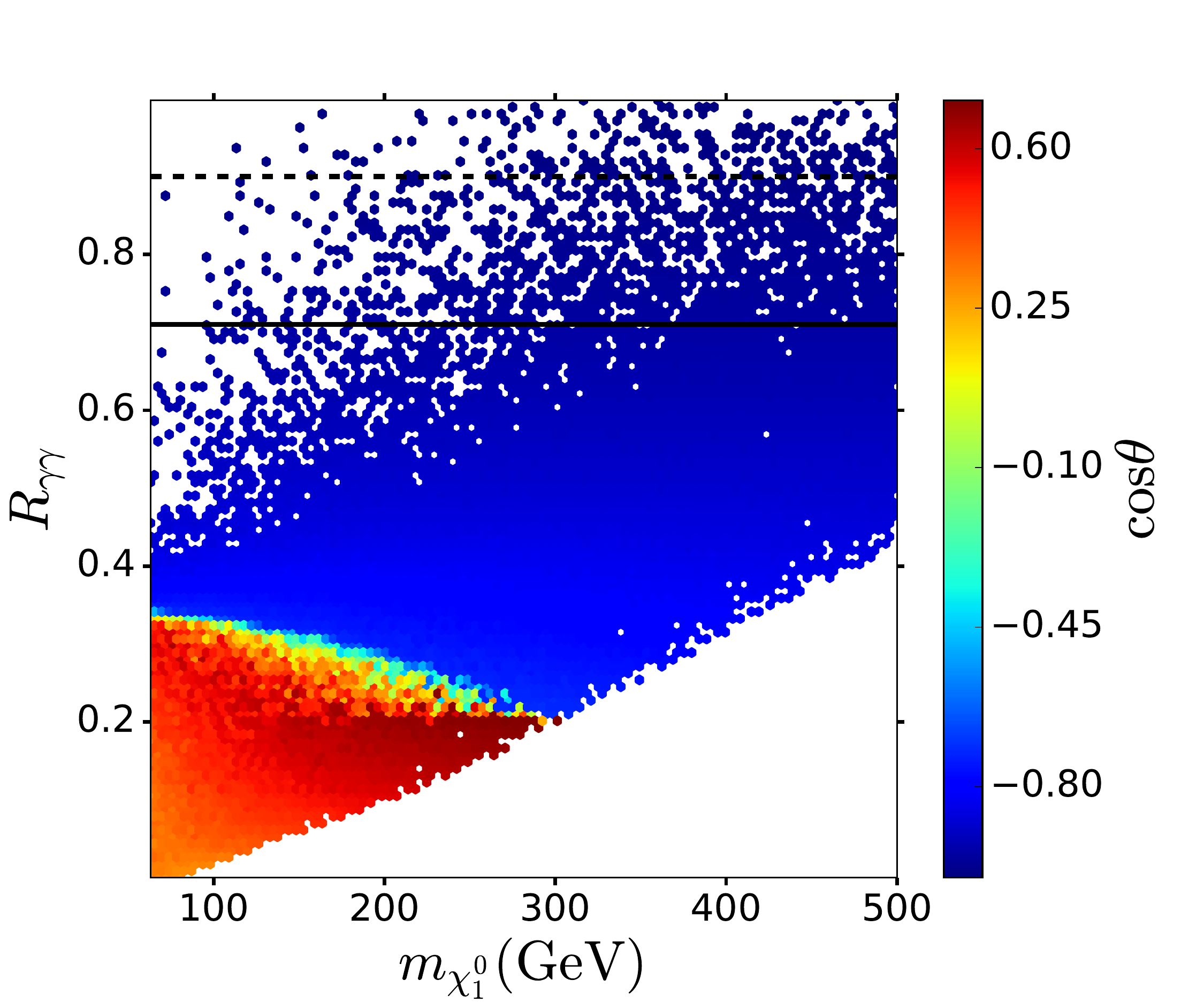}
\caption{Impact of the cosine of the mixing angle $\theta$ on the Higgs diphoton decay rate for the allowed values of the DM mass. The conventions are the same as those of Fig. \ref{fig:Rgg-1}.}
\label{fig:Rgg-2}
\end{center}
\end{figure}

We also consider the impact on the fermion mixing angle from $R_{\gamma \gamma}$  (see Fig \ref{fig:Rgg-2}).
For the region where $M_{\Sigma}<-m_{\chi^0_1}$ (left panel) the mixing angle must be small such that $|\cos \theta| \lesssim 0.3$ ($|\cos \theta|\lesssim 0.2$) in order to be consistent with ATLAS (CMS) measurements.
Accordingly, in that region the lightest charged fermion is mostly doublet.
On the other hand, the results for the region where $M_{\Sigma}>-m_{\chi^0_1}$ (right panel) point out that the lightest charged fermion is mostly triplet, with $|\cos \theta| \gtrsim 0.94$ ($|\cos \theta|\gtrsim 0.98$) for the ATLAS (CMS) data.
A comment regarding this region is in order: for $R_{\gamma \gamma} \sim 0.2$, the mixing angle exhibits a rather complex behavior which is seen as large changes in $\cos \theta$ (from -0.8 to 0.6) right next to a boundary where $\cos \theta \sim0.8$. 
This stems from the fact that at this boundary the triplet mass is changing sign, thus having an impact on the mixing angle behavior.

The results for the mixing angle in both region are important because they will have a direct impact on the production cross section of the heavier fermions at the LHC, as will be discussed below.

\subsection{Constraints from electroweak production searches}
Other LHC results that may potentially constraint the DTF model are those searching for electroweak production of neutralinos and charginos in different simplified SUSY models (with all other SUSY particles decoupled), where
the relevant detection channels are those with several leptons (and missing energy) in the final state. 
In the DTF,  $\chi_{1,2}^{\pm}$ and $\chi_{2,3}^{0}$ play the role of charginos and heavier neutralinos, respectively, with the same mass degeneracy that characterizes the simplified supersymmetric scenarios. 

The CMS collaboration has recently published results for such searches at $\sqrt{s}=13$ TeV and 35.9 fb$^{-1}$ \cite{Sirunyan:2018ubx}.
For the case of $m_{\chi_1^0}\lesssim500$ GeV and a nondegenerate spectrum,
the most sensitive channel is that with three final state leptons where at least two of them have opposite sign and same flavor. Thus, DM production proceeds via the following process
\begin{align}
  & q \bar{q}' \rightarrow W^{*\pm} \rightarrow \chi^\pm \chi_{2,3}^0:
\begin{cases}
  \chi^\pm \rightarrow \chi^0_{1} W^{*\pm} \rightarrow \chi_1^0 \ell^\pm \nu_\ell , \\
  \chi_{2,3}^0 \rightarrow \chi^0_{1} Z^* \rightarrow \chi_1^0 \ell^+ \ell^-.
  \end{cases}
\end{align}
where the mediators $\chi^{\pm}$ and $\chi^{0}$ are considered to be winos and thus mass degenerate, with the neutral fermion decaying 100$\%$ via $Z$ boson.
To recast the LHC constraints (and other experimental restrictions that will be discussed below) for the DTF we implemented the model in {\tt SARAH-4.12.3} package \cite{Staub:2013tta} whose output was used with {\tt SPheno-4.0.3} \cite{Porod:2003um} in order to obtain the particle spectrum and with {\tt MadGraph5\textunderscore}a{\tt MC@NLO} to obtain the production cross sections \cite{Alwall:2014hca}.

\begin{figure}[ht!]
\begin{center}
\includegraphics[scale=0.35]{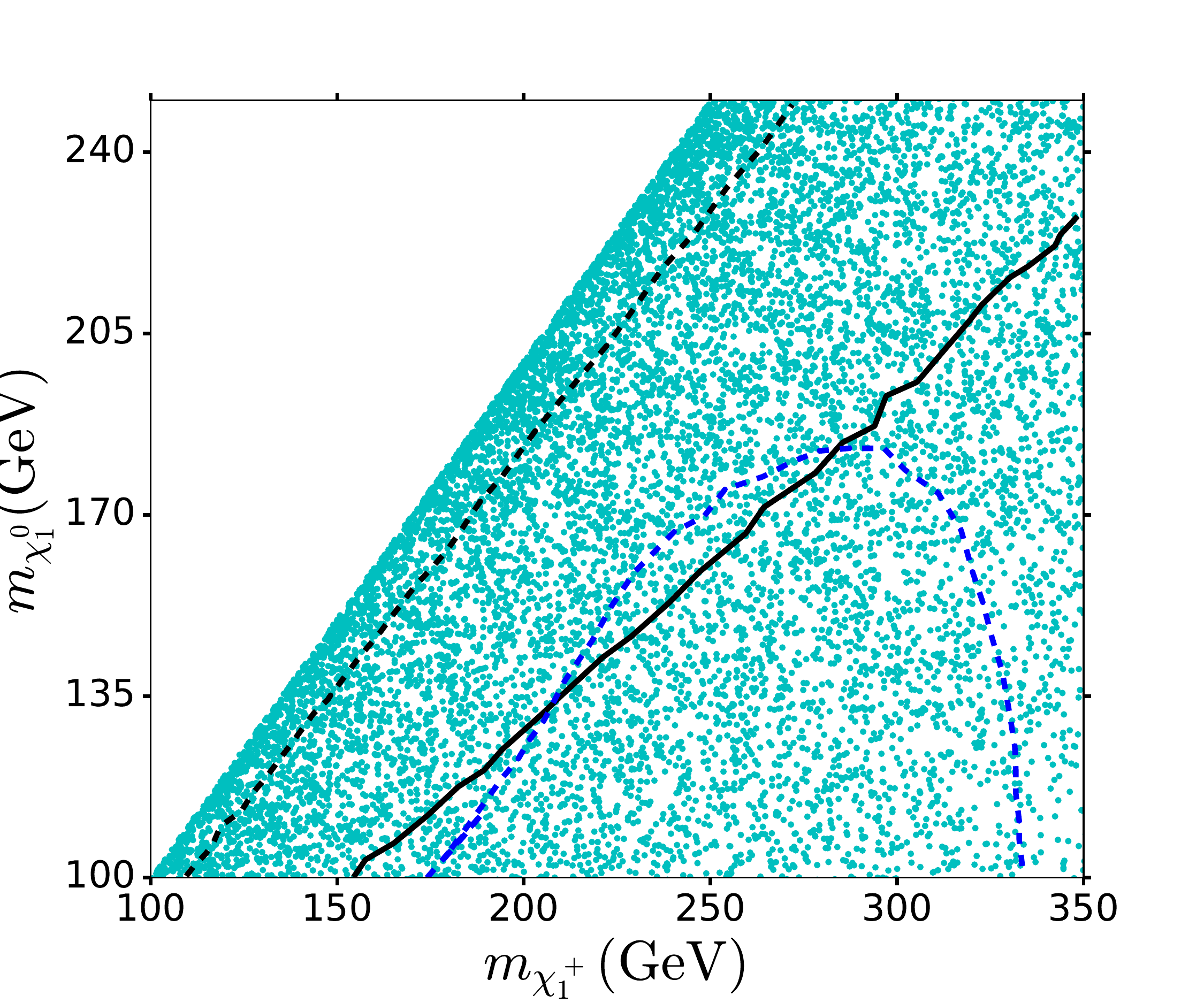}
\includegraphics[scale=0.35]{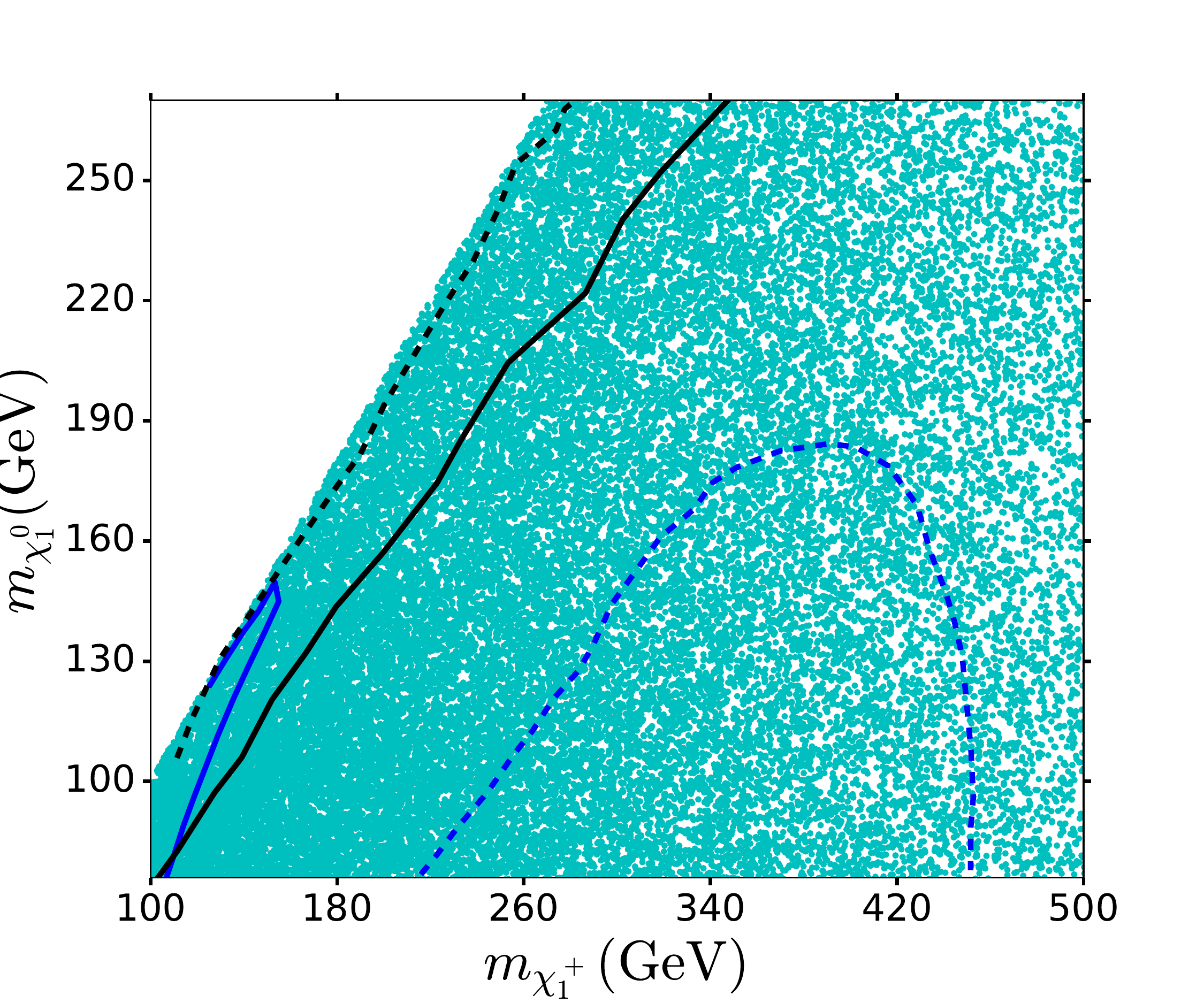}
\caption{DM mass versus lightest chargino mass for the regions where $M_{\Sigma}<-m_{\chi^0_1}$ (left panel)  and $M_{\Sigma}>-m_{\chi^0_1}$ (right panel). The region below the blue dashed line is excluded from CMS electroweak production while the region bounded by the blue solid line represents the exclusion by the ATLAS collaboration using compressed spectra. Points below the solid (dashed) black contour are excluded by the $R_{\gamma \gamma}$ results reported by the ATLAS (CMS) collaboration.}
\label{fig:prod}
\end{center}
\end{figure}

Fig. \ref{fig:prod} shows the constraints from electroweak production, where the excluded region corresponds to the points below the blue dashed line.
Moreover, the points below the solid and dashed black lines yield a lower diphoton decay ratio than the one allowed by ATLAS and CMS, respectively. In the region where $M_{\Sigma}>-m_{\chi^0_1}$ (right panel) the diphoton decay ratio restricts the lightest charged and the next-to-lightest neutral fermions to be mostly doublet. A consequence of this is that the production cross section is nearly the same even for all values of the allowed Yukawa coupling and the triplet mass, which means that the boundary of the excluded region is nearly independent of $y$ and $M_{\Sigma}$. Moreover, due to the mixing angle, the production cross section resembles that of SUSY Higgsino with all scalars decoupled. The figure shows that the strongest constraints come mostly from $R_{\gamma \gamma}$ , except for a small area where electroweak production cross section is more restrictive. Nonetheless, there are no additional restrictions placed on the free parameters $M_{\psi}$, $y$ and $M_{\Sigma}$.

In the region where $M_{\Sigma}<-m_{\chi^0_1}$, the diphoton decay rate restricts $\chi_2^0$ and $\chi_1^{\pm}$ to be mostly triplet. The production cross section is large but again independent of $y$ and $M_{\Sigma}$, and so, the region excluded by electroweak production is presented with only one contour. In this region, due to the larger production cross section, the curve is shifted to the left in the $m_{\chi_1^+}$ line and so $R_{\gamma \gamma}$ places the strongest constrains for the whole plain. 

\subsection{Constraints from compressed spectra searches}
The ATLAS collaboration has also published relevant results for the DTF for the case of compressed spectra \cite{Aaboud:2017leg}, {\it i.e.}, the next-to-lightest fermion is close in mass to the neutralino DM ($\leqslant$35 GeV) and a mass degeneracy between the next-to-lightest neutralino and lightest chargino. In that region, the DM production proceeds via:
\begin{align}
&q \bar{q}' \rightarrow W^{*\pm} \rightarrow \chi^\pm \chi_{2,3}^0:\,\,\, 
  \begin{cases}
    \chi^\pm \rightarrow \chi^0_{1} W^{*\pm} \rightarrow  \chi^0_{1} q \bar{q}' ,\\
    \chi_{2,3}^0 \rightarrow \chi^0_{1} Z^* \rightarrow \chi_1^0 \ell^+ \ell^- .
    \end{cases}
\end{align}
The search then focuses on two leptons with opposite sign and same flavor with soft momentum and large $\not\mathrel{E}$ which is present due to the two DM particles recoiling against initial state radiation.
For this search, small mass splittings are required, in order to ensure DM coannihilations.
In the DTF this low mass splitting is not needed, in fact $0 \lesssim m_{\chi_1^{\pm}}-m_{\chi_1^{0}} \lesssim 140$ GeV, however, we may still use the constraints for small mass splittings between the next-to-lightest $\chi^\pm$ and the DM.
We find that, for the region $M_{\Sigma}>-m_{\chi_1^0}$ where $\chi_1^{\pm}$ and $\chi_2^{0}$ are mostly triplet, and so the restrictions resemble those of the ATLAS collaboration which is shown with a solid blue contour. 
In terms of the free parameters, we find that the triplet mass is now restricted to be smaller than $\sim -165$ GeV, whereas the Yukawa coupling is not constrained.
For the case of $M_{\Sigma}<-m_{\chi_1^0}$, since $\chi_1^{\pm} \ \chi_2^{0}$ are mostly doublet, there is a lower production cross section and so the restriction is negligible.

\section{DM detection in a nonstandard cosmology}\label{NONSC}
In a nonstandard cosmology scenario, the late decay of a heavy scalar field could either increase or decrease the DM relic abundance compared to the standard calculation.
For the DTF\footnote{The case of the SUSY neutralino was studied in Ref.~\cite{Gelmini:2006pw}, whereas in Ref.~\cite{Cheung:2012qy} the wino-higgsino (similar to the DTF) case was also considered.}, it could be possible to satisfy the relic abundance, for instance,  due to the presence of a heavy scalar particle  which decays into heavier $Z_2$-odd particles that later on decay into the DM candidate, thus increasing the relic abundance compared to that of the model in the standard cosmology. 
Hence, we expect the DTF model to saturate, in one way or another, the DM relic abundance. Therefore, we look into current experimental constraints coming from direct searches and indirect detection via gamma rays. Since the diphoton decay is by far more restrictive than production at colliders, in this section we impose the $R_{\gamma \gamma}$ restriction coming from ATLAS and when relevant to the parameter space, we present the restriction arising in this observable from the CMS experiment.

\subsection{Direct detection}\label{DD}
\begin{figure}[t!]
\begin{center}
  \includegraphics[scale=0.5]{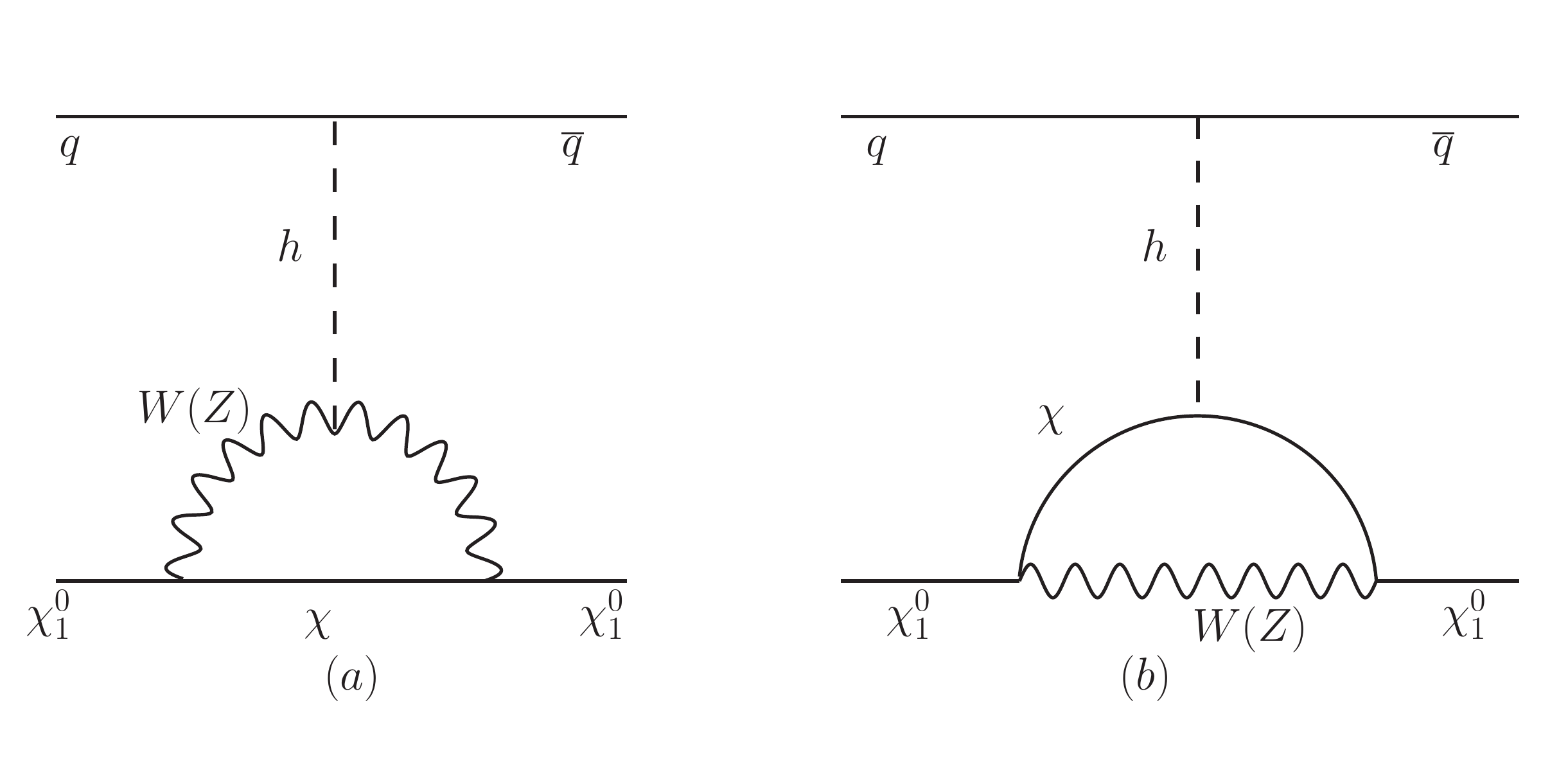}
\caption{Feynman diagrams contributing to the spin independent DD. The diagrams arise from a loop correction to the  $h\chi_1^0\chi_1^0$ coupling which is absent at tree-level when $\tan \beta=1$. The interaction shown in the left diagram is independent on the Yukawa coupling $y$ while the right one depends on it. The $Z_2$-odd $\chi$ fermion shown in both figures represent the charged $\chi_{1,2}^{\pm}$ (neutral $\chi_{2,3}^0$) fermion when the loop is mediated by $W$ ($Z$) boson. }
\label{fig:DD_h}
\end{center}
\end{figure}

Within the custodial limit scheme, the SI elastic scattering is only achieved at the loop level since both $g_{\chi_1^0\chi_1^0 Z}$ and $g_{\chi_1^0\chi_1^0 h}$ couplings vanish at tree-level.
However, at loop-level there are, in principle, several contributions that could be relevant.
First, there is an effective nonzero $\chi_1^0\chi_1^0h$ coupling originating from loops mediated by the new heavier fermions and weak gauge bosons, thus allowing for spin-independent direct detection (see Fig.~\ref{fig:DD_h}). 
Additionally, box diagrams mediated by gauge bosons and twist-2 operators \cite{Hisano:2011cs,Drees:1993bu} contribute to the SI cross section.
In principle, these two contributions should be taken into account to obtain a reliable calculation.
However, it has been shown that they are sub-leading except when the two contributions arising from the Higgs vertex corrections cancel each other out, which happens for low values of $\sigma_{SI}$ ($\lesssim 10^{-47}$cm$^{2}$) \cite{Drees:1993bu}.
Moreover, the authors of Ref.~\cite{Hisano:2011cs} have shown that when the cancellation happens, two-loop contribution to an effective scalar interaction with external gluons are of the same order as the box ones.
Since these calculations (boxes from gauge, twist-2 and two-loop) are quiet involved and only relevant for the case of specific cancellations, we will not take them into account for the calculation of the SI cross section. Moreover, they tend to create a larger suppression of the cross section that is already out of reach of current experiments. As a result, the restrictions that we will present below from DD are not strongly affected by this assumption.

In order to obtain the most up-to-date limits from DD, we calculated the effective $g_{h\chi_1^0\chi_1^0}$ coupling following  Ref.~\cite{Freitas:2015hsa} and used that to compute the spin independent (SI) cross sections. We then compared to the current upper limits on the DM-nucleon SI scattering cross section, where the strongest ones (within the DM mass range we are considering) are those reported by the XENON1T collaboration \cite{Aprile:2018dbl}. We also show the projected sensitivity of DARWIN \cite{Aalbers:2016jon}, the most sensitive DD experiment planned for DM at the electroweak scale. However, the expected SI cross section around the DARWIN limit must be taken with a grain of salt since sub-leading corrections might change $\sigma_{SI}$ in that region. 

%%%%%%%%%%%%%%%%%%%%%%%%%%
%%% SUPERIOR %%%%%%%%%%%%%
%%%%%%%%%%%%%%%%%%%%%%%%%%

\begin{figure}[t!]
\begin{center}
\includegraphics[scale=0.35]{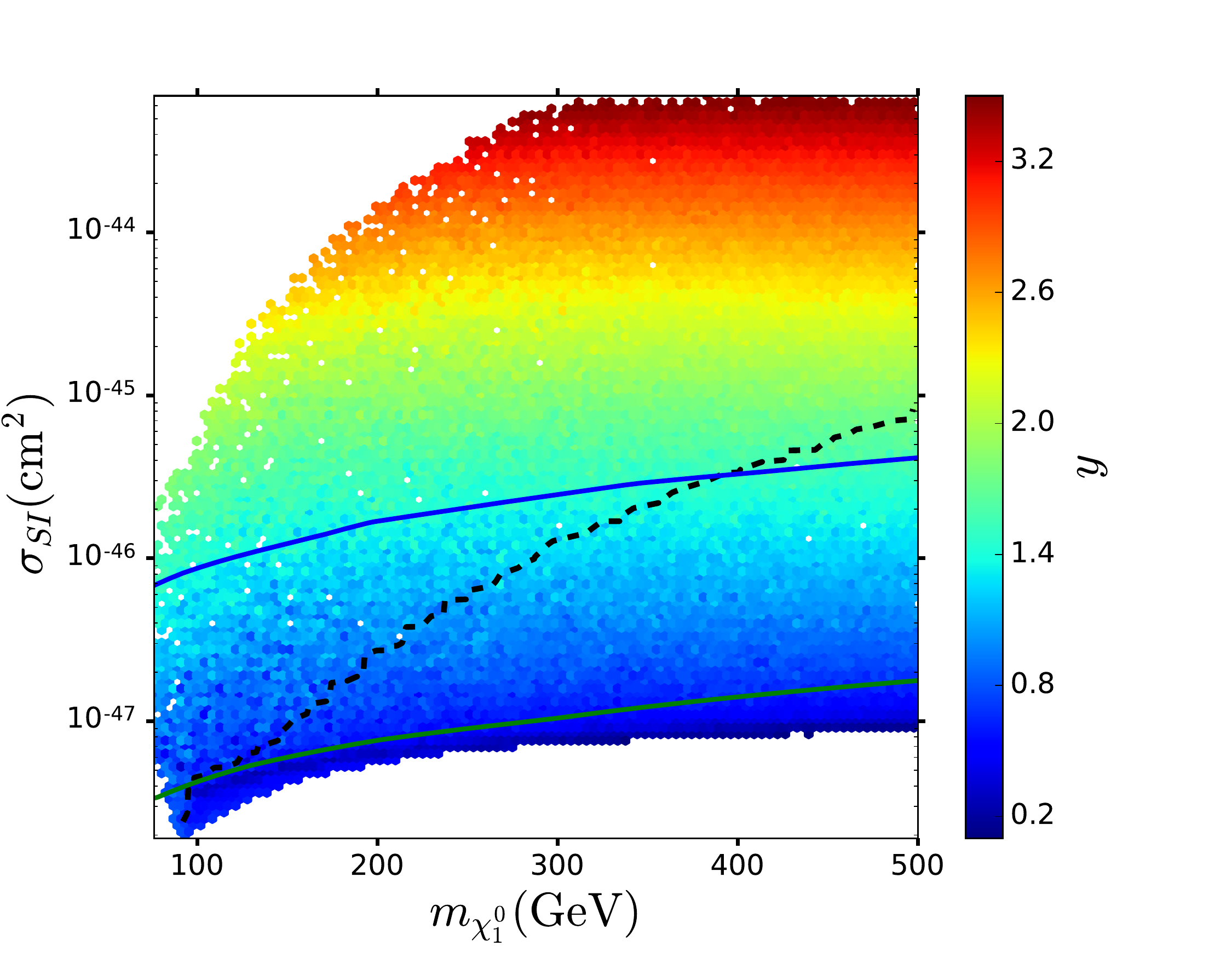}
\includegraphics[scale=0.35]{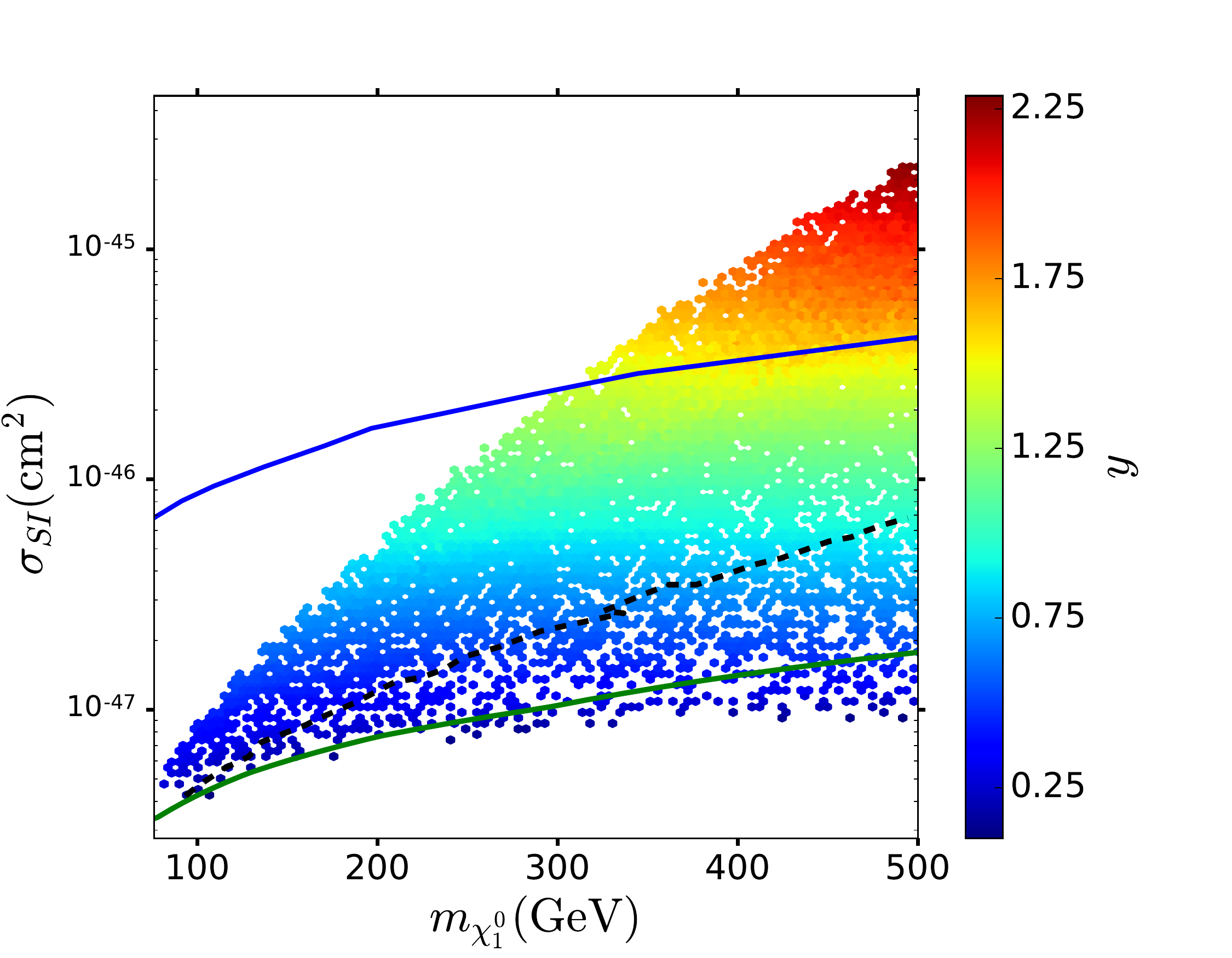}
\caption{Spin-independent cross section for the regions $M_{\Sigma}<-m_{\chi_1^0}$ (left) and $M_{\Sigma}>-m_{\chi_1^0}$ (right). The blue curve represents the upper limit imposed by XENON1T \cite{Aprile:2018dbl} whislt the green curve shows the projected sensitivity of DARWIN \cite{Aalbers:2016jon}.
The black dashed line represents the limit when the $R_{\gamma \gamma}$ restriction from CMS is considered.}
\label{fig:Un-sup-DD}
\end{center}
\end{figure}

In Fig. \ref{fig:Un-sup-DD} we display the results for the spin-independent cross section as a function of the DM mass,  for the regions $M_{\Sigma}<-m_{\chi_1^0}$ (left) and $M_{\Sigma}>-m_{\chi_1^0}$ (right).
It follows that XENON1T restricts the coupling to be less than 1.75 if the lower bound on $R_{\gamma \gamma}$ from ATLAS is imposed.
The dashed black line in both panels shows the CMS lower bound on $R_{\gamma \gamma}$, which excludes models even further and for the region $M_{\Sigma}>-m_{\chi_1^0}$ imposes $y \leq 1.2$.
We also checked the impact of DD results on the other free parameter of the model, $M_{\Sigma}$, but we found that they place no further restrictions on it.
The prospects coming from the DARWIN experiment correspond to the green solid line, which show that couplings as small as 0.5 may be probed. It is worth mentioning that the lower limit on the SI cross section is due to the cancellation between the two one-loop corrections to the $h\chi_1^0\chi_1^0$ vertex; hence, in order to have a precise value of $\sigma_{SI}$ in this region, a more detailed calculation is necessary.

\subsection{Indirect detection from dwarf spheroidal galaxies}\label{ID1}
\begin{figure}[t!]
\begin{center}
\includegraphics[scale=0.35]{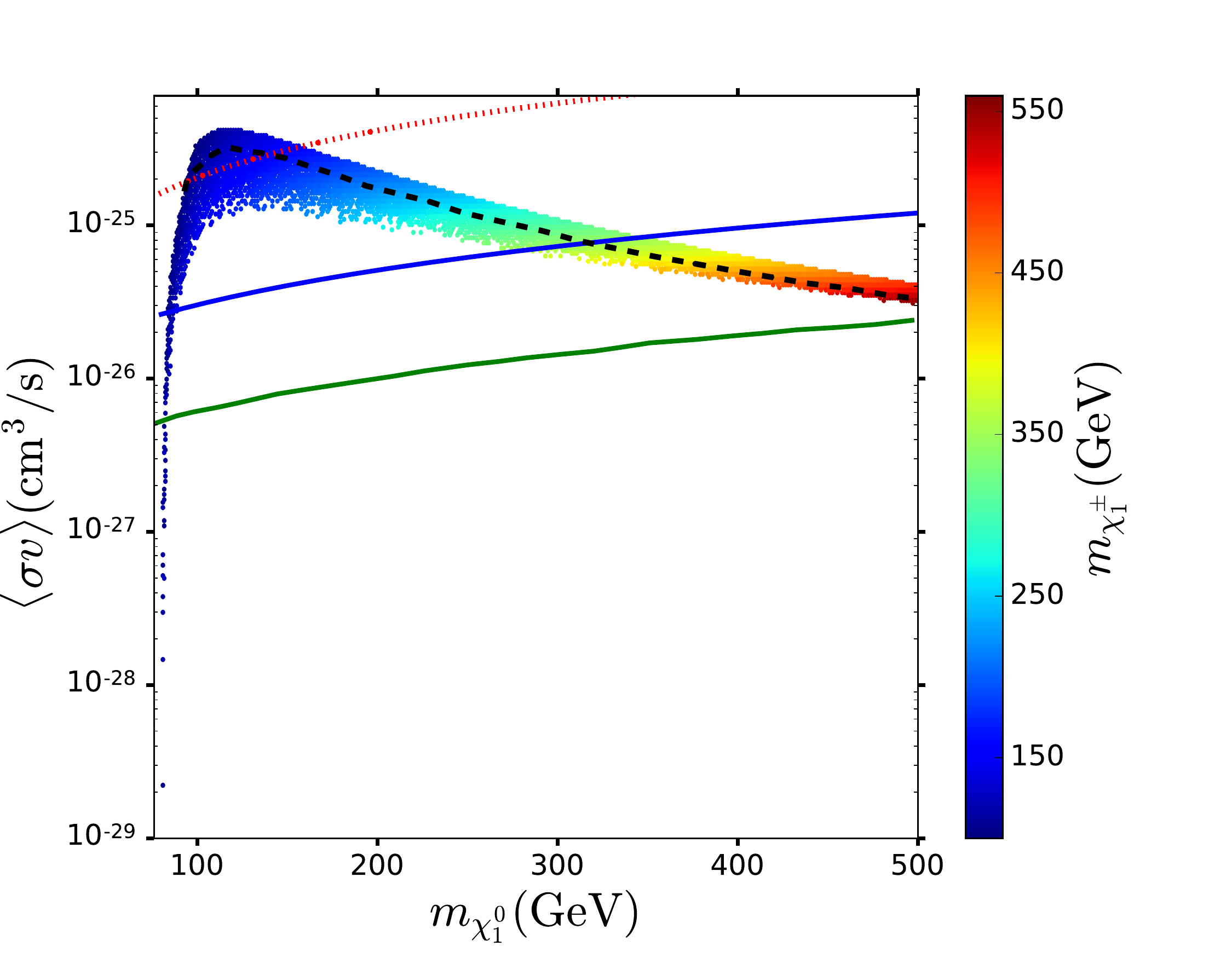}
\includegraphics[scale=0.35]{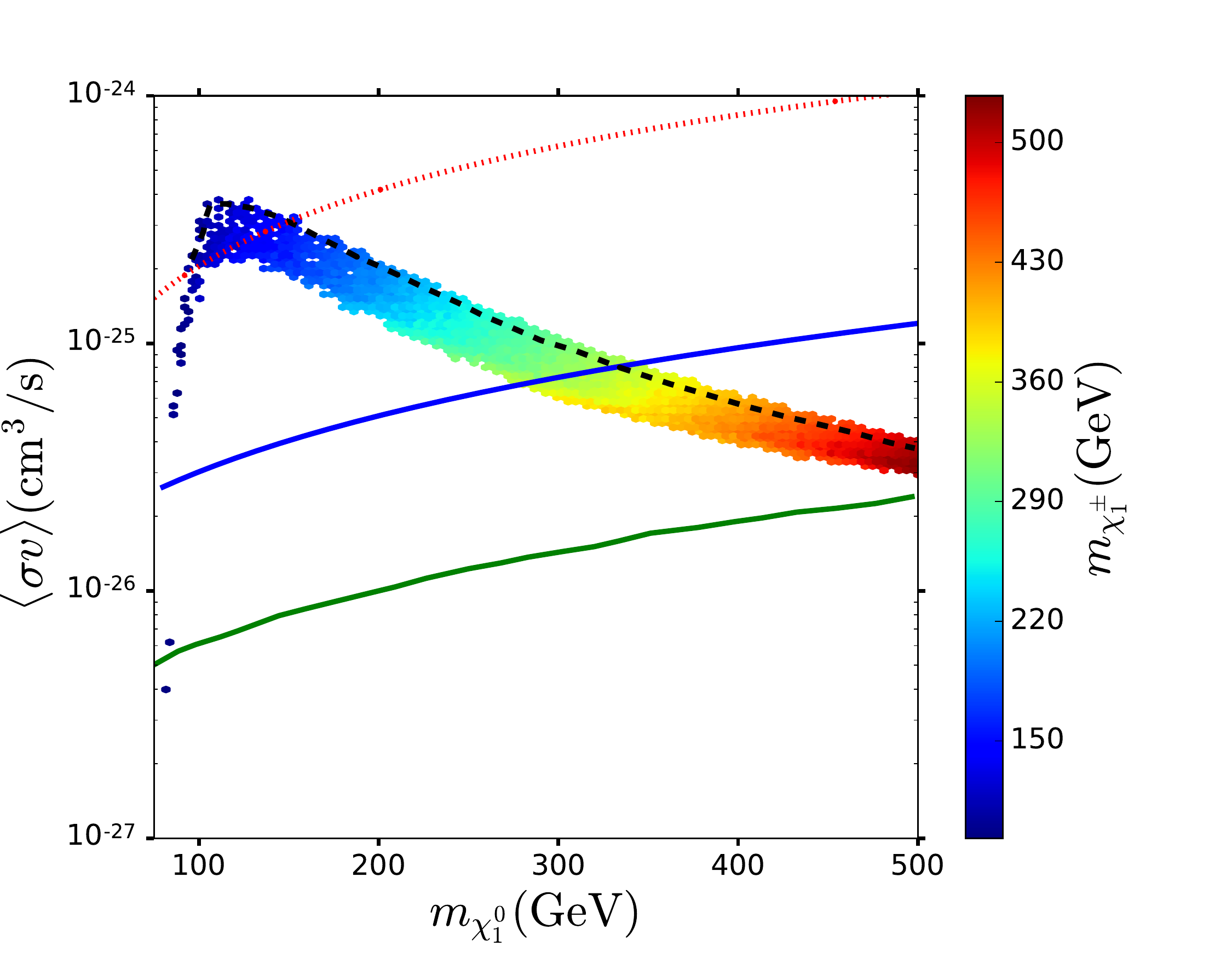}
\caption{ID restrictions and prospects coming from the observation of dSphs of the Fermi satellite applied to the regions $M_{\Sigma}<-m_{\chi_1^0}$ (left panel) and $M_{\Sigma}>-m_{\chi_1^0}$ (right panel). The blue and green curve show current limits from the $W^+W^-$ channel for 6 years of observation and 15 dSphs, and the projected sensitivity for 45 dSphs and 15 years of observation, respectively.
  Whereas points above the red dotted line are excluded from CMB measurements by the Planck Collaboration.
  Points below the black dashed line are excluded when the $R_{\gamma \gamma}$ restriction from CMS is considered.}
\label{fig:sup-inf-ID}
\end{center}
\end{figure}

In regions of high dark matter density such as dwarf spheroidal galaxies (dSphs) or the center of the Milky Way, DM particles may more easily find each other and annihilate into SM particles.
The dSphs are particularly interesting because of their proximity to the Milky Way, their high DM to baryon mass-ratio, and their low background, thus making the DM detection via gamma-rays feasible. 
The Fermi satellite has searched for gamma rays in dSphs founding no deviations from the expected spectrum, which has lead to upper limits on the thermally averaged DM annihilation cross section \cite{Ackermann:2015zua}.
It is worth noting that in this model DM annihilation can also affect the cosmic microwave radiation (CMB). If DM annihilates during the time of recombination, it will inject energy that will ionize Hydrogen. This will have a direct effect on the anisotropies of the CMB, thus, altering what is currently observed. Therefore, measurements of the CMB can constrain the parameter space of DM models, with the advantage that for this observable astrophysical uncertainties do not play a role \cite{Kawasaki:2015peu,Lopez-Honorez:2013lcm}.

For the DTF, the DM annihilation proceeds in the same channels as the ones in the early Universe, {\it i.e.}, via $t$- and $u$-channel annihilation into $W^+ W^-$ and $Z Z$ bosons. The gauge bosons then decay and produce, for instance, gamma rays that may be detected as an excess in the spectrum.To obtain the constraints we calculated the thermally averaged cross section using the public available package {\tt micrOMEGAS} \cite{Belanger:2013oya} and used this to compare with limits reported in Ref.~\cite{Ackermann:2015zua}.
In the DTF this cross section is, to a leading order approximation, independent of the DM velocity. Thus, its value matches that of the early universe, which allow us to compare our results with the limits reported in Ref.~\cite{Aghanim:2018eyx}.

The results are shown in Fig.~\ref{fig:sup-inf-ID} where all points shown satisfy the ATLAS $R_{\gamma \gamma}$ constraint and DD bounds as explained in previous sections. As can be seen, the Fermi satellite observation over 15 dSphs imposes stringent limits on the model in a such a way that a large portion of the DM mass range is ruled out. Moreover, stringent limits on the mass of the next-to-lightest fermion also arise, since such particles act as the mediators in the $t$- and $u$-channels of the DM annihilation.
On the other hand, though CMB measurements do place constraints, they are far less restrictive than those coming from dSphs. 
For the region where $M_{\Sigma}<-m_{\chi_1^0}$ we find that $86\,\text{GeV}<m_{\chi_1^0}<280$ GeV is already ruled out, this also leads to a restriction on $m_{\chi^{+}}>340$ GeV for $m_{\chi_1^0}>280$ GeV. On the other hand, for the region where $M_{\Sigma}>-m_{\chi_1^0}$ we find that the diffuse spectrum requires that $m_{\chi_1^0}>280$ GeV, $m_{\chi^{+}}>300$ GeV while $M_{\Sigma}\lesssim -230$ GeV.
The expected 15 years and 45 dSphs observation will explore the whole region of the right panel and will leave a very narrow range of $m_{\chi_1^0}$ of $\sim$ 80 GeV un-explored.
We also note that points that satisfy the $R_{\gamma \gamma}$ restriction of the CMS experiment are those with the higher $\langle \sigma v \rangle$ since both observables depende on the mixing angle and are maximal for large $y$\footnote{For $R_{\gamma \gamma}$ the dependence was already shown in Sec. \ref{collider}, while for the diffuse spectrum, the dependence on $\cos\theta$ enters through the vertices of the annihilation channels. In Appendix \ref{AppendixA} this dependence is shown for the DM interaction with the $W^{\pm}$ gauge boson and a $Z_2$-odd charged fermion.}.

\subsection{Indirect detection from gamma-ray lines }\label{ID2}
Another promising detection channel is DM annihilation into two photons within regions with high DM density. 
In this case, the photon energies will be closely related to the DM mass leading to a spectrum exhibiting a sharp peak referred to as a linelike feature \cite{Funk:2013gxa}. 
In this regard, the Fermi~\cite{Ackermann:2015lka} and H.E.S.S. \cite{Abdalla:2016olq,Abdallah:2018qtu} collaborations have looked for gamma-ray lines coming from the center of the Milky Way,  with no evidence of DM so far. This in turn has lead to constraints on the DM $ \langle \sigma v \rangle_{\gamma\gamma}$ annihilation into photons.

In the DTF, the DM annihilation into two photons is mediated by heavier $Z_2$-odd fermions interacting with vector and Goldstone bosons. Though the annihilation cross section in this case is loop suppressed, it may be possible to place constraints. In order to calculate the $ \langle \sigma v \rangle_{\gamma\gamma}$ we follow the procedure given in Ref.~\cite{Garcia-Cely:2016hsk} (the specific calculations along with the topologies that contribute are given in Appendix \ref{AppendixA}).
After considering all the restrictions coming from collider, DD and ID in the diffuse spectrum, our results show that the Fermi and H.E.S.S. results do not place additional constraints on the model for both $M_{\Sigma}<-m_{\chi_1^0}$ and $M_{\Sigma}>-m_{\chi_1^0}$ regions since $ \langle \sigma v \rangle_{\gamma\gamma} \sim 10^{-29}\, \text{cm}^3/\text{s}$, which is nearly an order of magnitude lower than the most sensitive results which are presented by the H.E.S.S collaboration in Ref.~\cite{Abdallah:2018qtu}. As a result, this observable does not restrict the parameter space of the model.

\section{DM detection in multicomponent dark sectors}\label{multiDM}
An interesting possibility that has recently taken momentum is for the DM to be composed of different sectors, which is a far more general setting than the usual one DM candidate. For instance, the observed relic density could be the result of WIMP and Axion particles. In this case, it is possible that the sectors do not communicate, and so they behave as two completely independent DM particles, without affecting each other's relic density and experimental bounds. For this section we will consider the WIMP DM candidate from the DTF to be part of multicomponent DM, that is, we obtain experimental bounds for models where the WIMP's relic density is less than or equal to the central value reported by the Planck collaboration, $\Omega_{\text{Planck}}$ \cite{Aghanim:2018eyx}.

\begin{figure}[ht!]
\begin{center}
\includegraphics[scale=0.35]{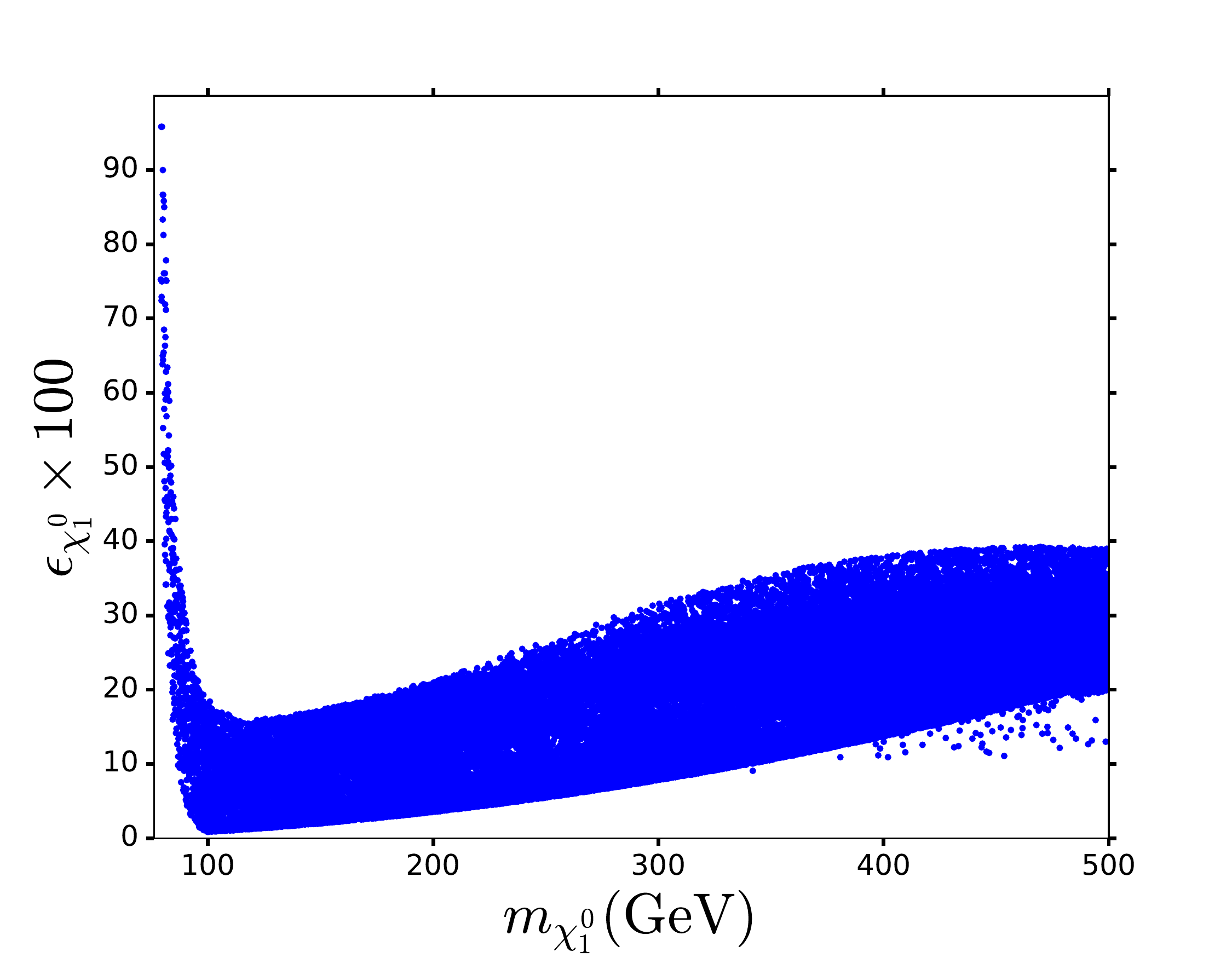}
\includegraphics[scale=0.35]{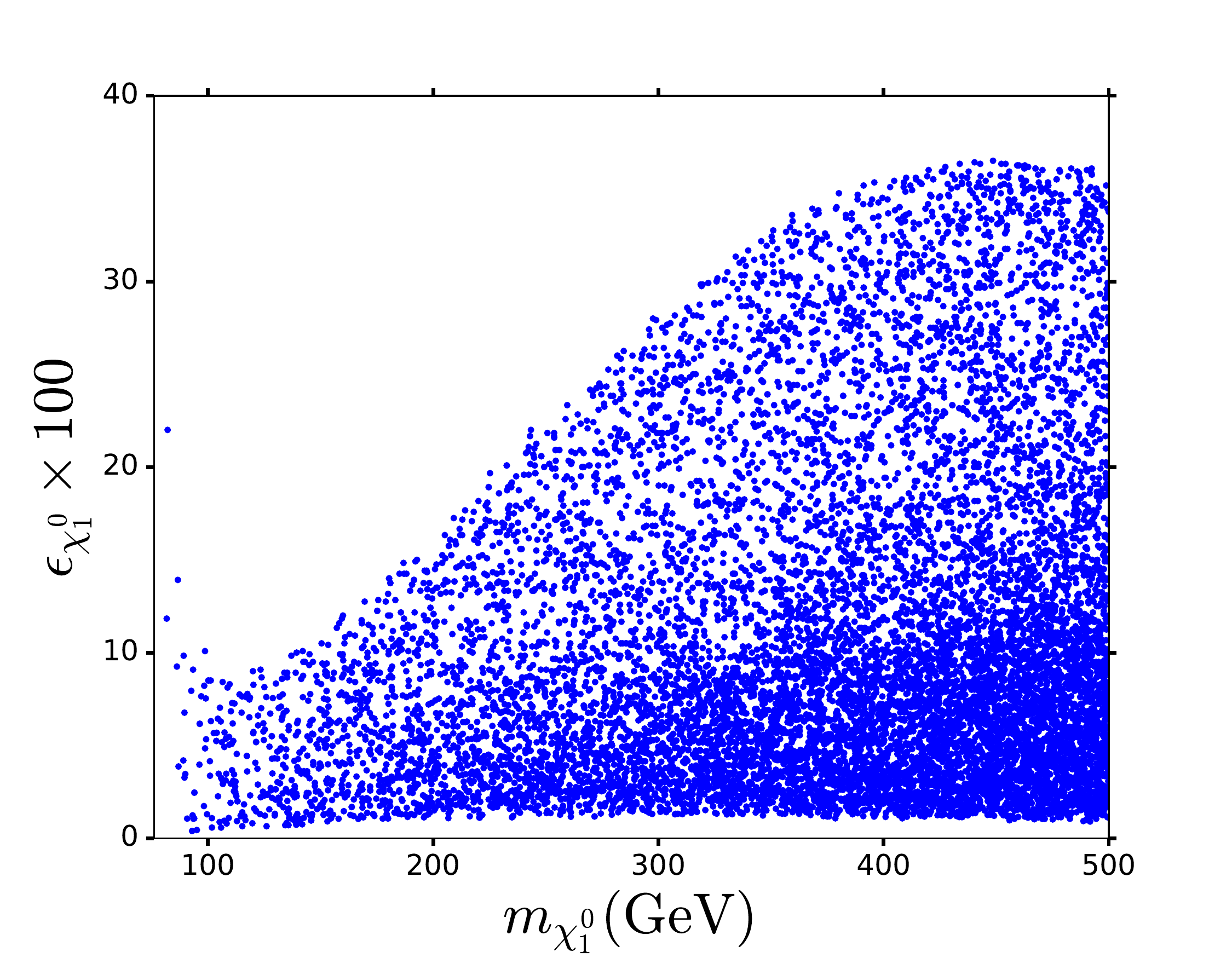}
\caption{  $\epsilon_{\chi_1^0}$ vs. $m_{\chi_1^0}$ for $M_{\Sigma}<-m_{\chi_1^0}$ (left panel) and $M_{\Sigma}>-m_{\chi_1^0}$ (right panel). All points satisfy collider bounds presented in Sec. \ref{collider}.}
\label{fig:perc_omega}
\end{center}
\end{figure}

Figure \ref{fig:perc_omega} shows the ration $\epsilon_{\chi_1^0}=\Omega_{\chi_1^0}/\Omega_{\text{Planck}}$ as a function of $m_{\chi_1^0}$. 
For the region where $M_{\Sigma}<-m_{\chi_1^0}$, the relic abundance is at most 40$\%$ of the observed value except for the narrow region where $m_{\chi_1^0} \sim 80$ GeV (where annihilation into weak gauge bosons is kinematically suppressed).
On the other hand, in the region where $M_{\Sigma}>-m_{\chi_1^0}$ there are no models that saturate relic density, and so, the DTF accounts at most 40$\%$ of the Universe DM content.
We must add a comment here, unlike the previous section, we are assuming that the DM arises from a standard cosmology scenario, and in that sence the relic abundance of the WIMP DM is the one calculated through the usual method of solving the Boltzman's equation (the one calculated via {\tt micrOMEGAS} \cite{Belanger:2013oya}).

\begin{figure}[ht!]
\begin{center}
\includegraphics[scale=0.35]{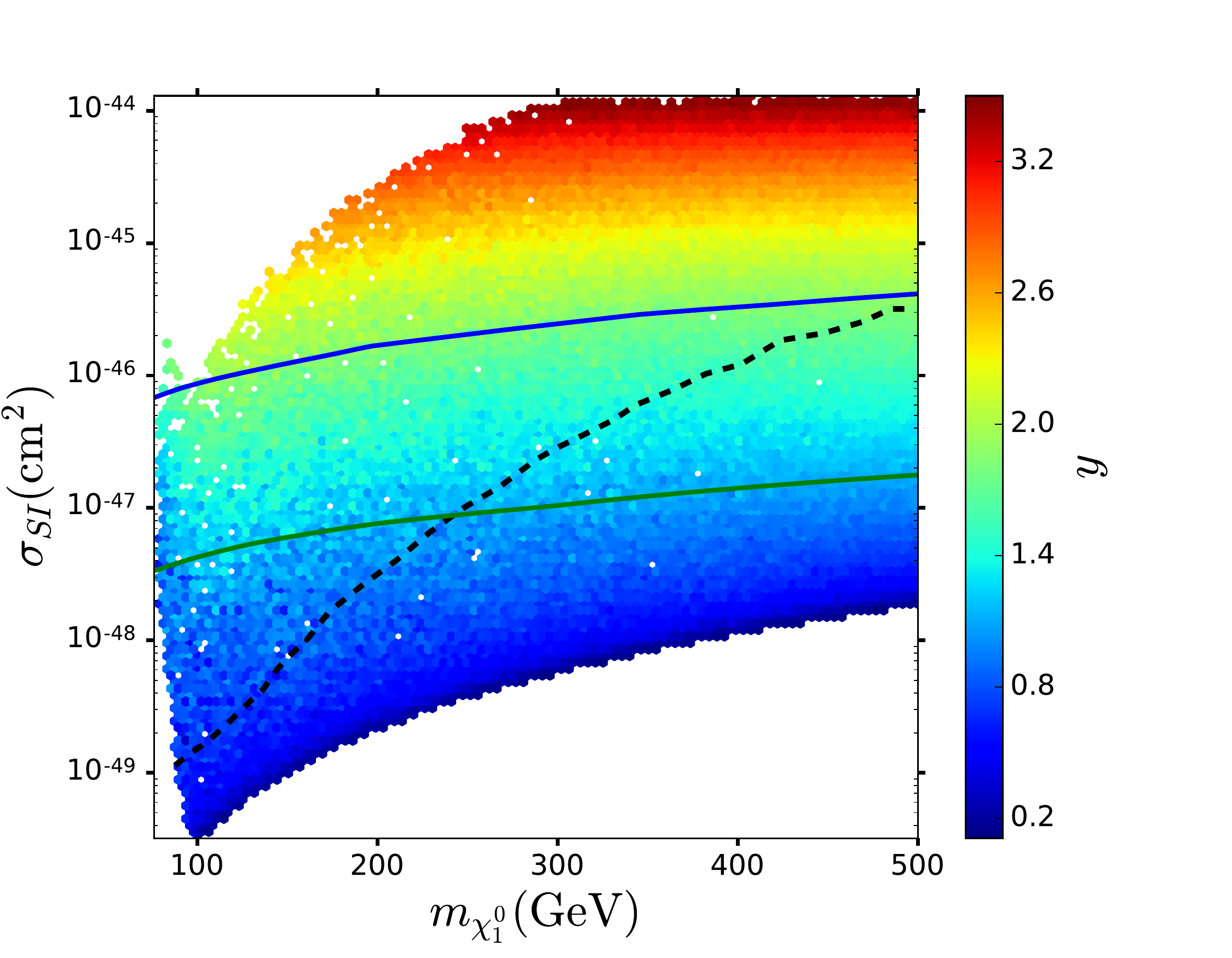}
\includegraphics[scale=0.35]{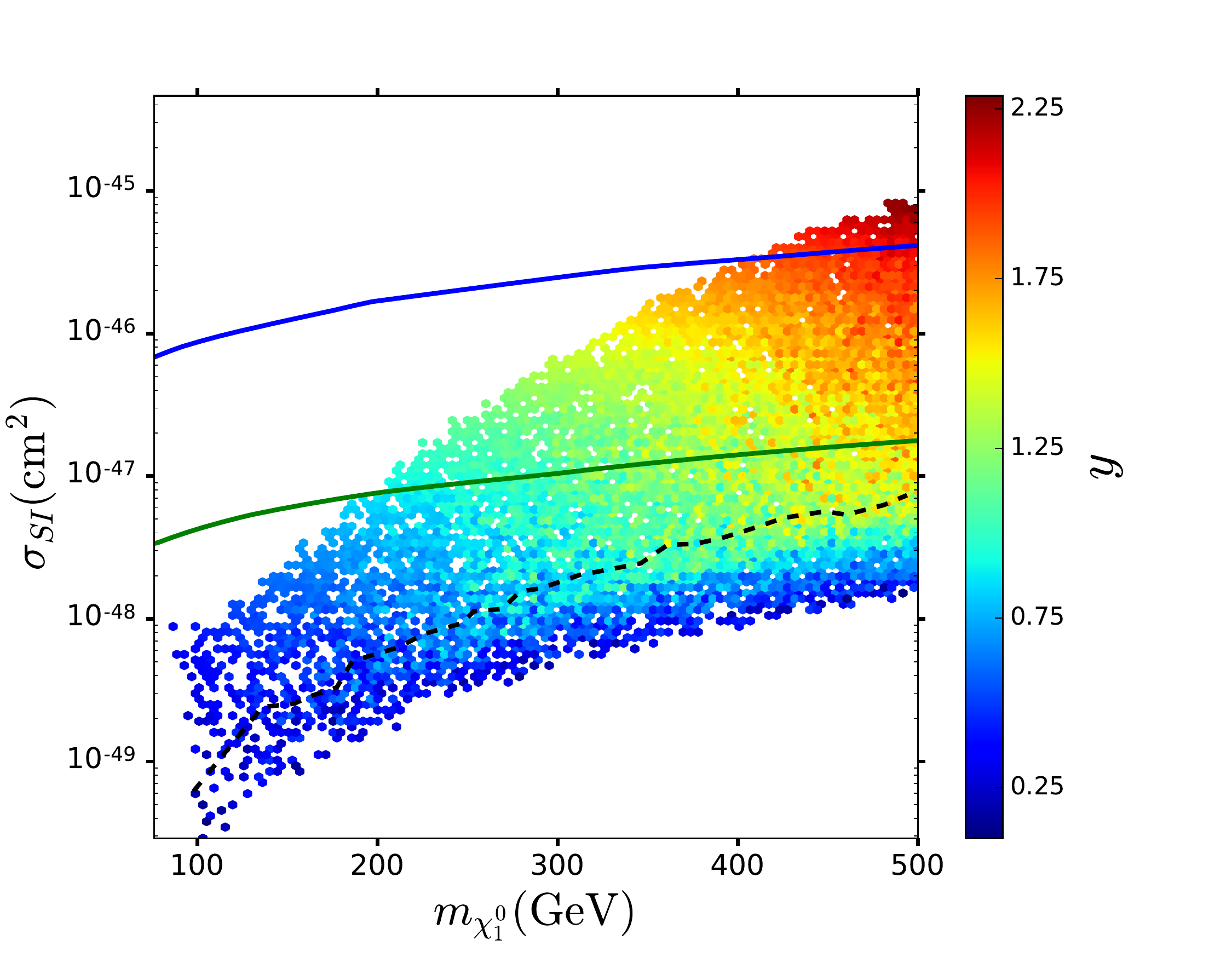}
\caption{Direct detection results for  $M_{\Sigma}<-m_{\chi_1^0}$ (left panel) and $M_{\Sigma}>-m_{\chi_1^0}$ (right panel), the conventions are the same as in Fig.~\ref{fig:Un-sup-DD}.}
\label{fig:sup-inf-DDre}
\end{center}
\end{figure}

Now we set out to investigate experimental bounds on the model. For colliders, the restrictions are the same as those presented in Sec. \ref{collider} since they are independent of the DM abundance. On the other hand, DD and ID rates do depend in the local DM density, and as a result the constraints presented in Sec.~\ref{NONSC} will be different in this scenario.
%To quantify this, it is useful to define the parameter $\epsilon_{\chi_1^0}=\frac{\Omega_{\chi_1^0}h^2}{\Omega_{CDM}h^2}$ \cite{Cao:2007fy,Hur:2007ur} which will re-scale the WIMP DM abundance compared to the observed relic density.
To quantify this, we used the parameter $\epsilon_{\chi_1^0}$ \cite{Cao:2007fy,Hur:2007ur} to re-scale the DD and ID observables. 
For the case of DD, the expected scattering rate will be rescaled by $\epsilon_{\chi_1^0}$ which means that the SI cross section is effectively rescaled to be $\sigma_{SI}=\epsilon_{\chi_1^0}\sigma_{SI-\chi_1^0}$; hence, DD constrains are now relaxed. The results are shown in Fig.~\ref{fig:sup-inf-DDre} for $M_{\Sigma}<-m_{\chi_1^0}$ (left panel) and $M_{\Sigma}>-m_{\chi_1^0}$ (right panel). The left panel shows that for models that satisfy the lowest ATLAS limit on $R_{\gamma \gamma}$, DD imposes $y \leqslant 2.1$ while for models that satisfy lowest CMS limits $y \leqslant 1.9$. On the other hand, in the right panel, for models that satisfy the lowest ATLAS limit on $R_{\gamma \gamma}$, DD imposes $y \leqslant 2.2$ while for models that satisfy lowest CMS limits $y \leqslant 0.95$ which means that in this case CMS diphoton decay is more restrictive than DD (even considering DARWIN prospects).  

For indirect detection, the situation is far less restrictive because the thermally averaged cross section is rescaled by a factor of $\epsilon_{\chi_1^0}^2$, thus suppressing it.
% The exact same situation occurs for the mono-energetic gamma rays.
As a result, ID does not impose additional constraints on the model.

\section{Conclusions}\label{sec:conc}
In this work we have studied a simplified DM model, the doublet-triplet fermionic model, where the SM is enlarged with an $SU(2)_L$ vectorlike doublet and a Majorana triplet, both new fields are odd under a $Z_2$ symmetry while the SM fields are even. As a result the new fields Lagrangian include two Yukawa type interactions with the Higgs field.
It follows that when the two allowed Yukawa couplings of the dark sector are equal, the model exhibits a custodial symmetry and, when the DM particle is pure doublet, the diagonal couplings of it with the Higgs and $Z$ bosons are  forbidden at tree-level.
In this case, the model may saturate the relic abundance at the electroweak scale, but that comes with the disadvantage of a strong suppression of the Higgs diphoton decay rate.

For this reason, we have considered the model (in the aforementioned case) framed within two different scenarios: one where the relic abundance is set by nonstandard cosmology, {\it i.e.}, we assumed the relic abundance is saturated due to new physics before BBN, and another where the DM is made up of multiple particles which do not affect each other's abundance or DD limits.
As a result, the mass of the heavier charged and neutral fermions may lie close to the DM mass, which lifts partly the $R_{\gamma \gamma}$ restriction.

Regarding DD and ID for the nonstandard cosmology scenario, we found that Xenon1T results demand a Yukawa coupling $y<1.75$,  whereas the Fermi results imply that the DM mass is in general restricted to be $m_{\chi_1^0} < 280$ GeV except for a narrow region of $m_{\chi_1^0} \sim 80$ GeV when $M_{\Sigma} < m_{\chi_1^0}$.
On the other hand, for the scenario of the DM as part of the multicomponent dark sectors we found that DD impose a less severe constraint on the Yukawa coupling ($y<2.2$) while current ID does not impose additional constraints on the model.  

\section*{Acknowledgements}
We are thankful to Diego Restrepo, Andrés Rivera and Guillermo Palacio for enlightening discussions. We are also thankful to Florian Staub for help with the SARAH package and Olivier Mattelaer for help with MadGraph.
A. B. has been supported by Colciencias, Universidad EIA and Fulbright Colombia.
O.Z. has been supported by the Sostenibilidad program of Universidad de Antioquia UdeA  and by COLCIENCIAS through the grants 111565842691 and 111577657253. O.Z. acknowledges the ICTP Simons associates program and the kind hospitality of the Abdus Salam ICTP where the final stage of this work was done. 
\appendix{}

\section{Annihilation into two photons}\label{AppendixA}
In this appendix, we give the explicit calculation to obtain the thermally averaged cross section for the annihilation of DM in the DTF model into two photons (e.g. gamma-ray lines). The procedure was obtained with the results presented in Ref. \cite{Garcia-Cely:2015khw}.

\begin{figure}[t!]
\begin{center}
\includegraphics[scale=0.3]{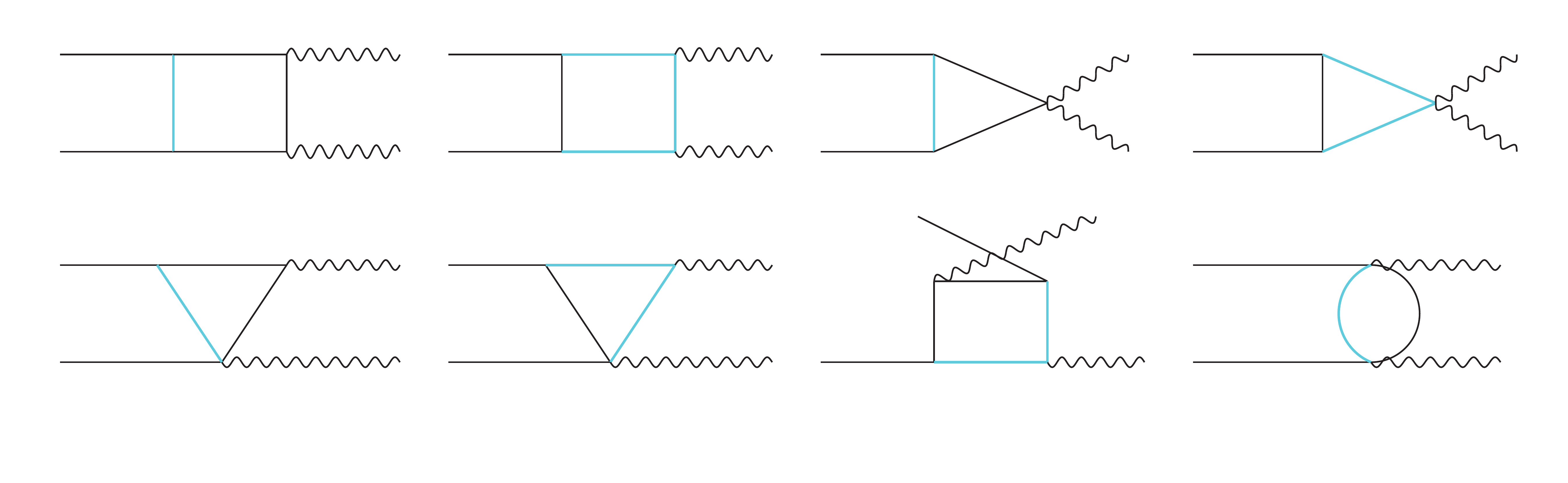}
\caption{Topologies that lead to the annihilation of DM into two photons. The external straight lines represent the DM particles, whereas the internal ones represent a charged $Z_2$ odd fermion (shown with a cyan solid line), gauge boson or Goldstone boson (shown with a black solid line). The external wavy lines represent the photons (the gamma-rays).}
\label{fig:gamma-lines}
\end{center}
\end{figure}

The thermally averaged cross section is given by
\begin{align}
\langle \sigma v \rangle =\frac{1}{4} \frac{|\mathcal{B}|^2}{32 \pi m_{\chi_1^0}^2},
\end{align}
where $\mathcal{B}=\mathcal{B}_W+\mathcal{B}_S$. Here $\mathcal{B}_W$ and $\mathcal{B}_S$ denote the contributions coming from the charged gauge bosons ($W$) and scalars (Goldstone bosons, $S$) running in the loop, respectively (see Fig.~\ref{fig:gamma-lines}).
They read
\begin{align}
\mathcal{B}_i=& \frac{\alpha}{\pi} \bigg( x_1 \frac{C_0 (0,1,-1,r_{\mathrm{even}}^2,r_{\mathrm{even}}^2,r_{\mathrm{odd}}^2)}{(r_{\mathrm{odd}}^2-r_{\mathrm{even}}^2)(1+r_{\mathrm{odd}}^2-r_{\mathrm{even}}^2)} + x_2 \frac{C_0 (0,1,-1,r_{\mathrm{odd}}^2,r_{\mathrm{odd}}^2,r_{\mathrm{even}}^2)}{(r_{\mathrm{even}}^2-r_{\mathrm{odd}}^2)(1-r_{\mathrm{odd}}^2+r_{\mathrm{even}}^2)} \\ \nonumber 
& + x_3 \frac{C_0 (0,4,0,r_{\mathrm{even}}^2,r_{\mathrm{even}}^2,r_{\mathrm{even}}^2)}{(1+r_{\mathrm{odd}}^2-r_{\mathrm{even}}^2)} + x_4 \frac{C_0 (0,4,0,r_{\mathrm{odd}}^2,r_{\mathrm{odd}}^2,r_{\mathrm{odd}}^2)}{(1-r_{\mathrm{odd}}^2+r_{\mathrm{even}}^2)} \bigg).
\end{align}
Here $r_{\mathrm{even (odd)}}=m_{even (odd)}/m_{\chi_1^0}$ with the label even (odd) indicating that the particle is $Z_2$ even (odd) and $C_0 (r_1^2,r_2^2,r_3^2,r_4^2,r_5^2,r_6^2)$ is the usual Passarino-Veltman function \cite{Peskin:1995ev}. In the case of the charged Goldstone boson the mass $m_{even}=m_{W}$. On the other hand, the factors $x_i$ are different depending if the mediator is a scalar or a vector boson:

\textbf{Scalars}
\begin{align}
x_1=& \sqrt{2} r_{\mathrm{even}}^2 \big(r_{\mathrm{even}}^2-r_{\mathrm{odd}}^2-1 \big)(g_{Ls}^2+g_{Rs}^2), \\ \nonumber
x_2=& \sqrt{2} r_{\mathrm{even}}^2 \big(r_{\mathrm{even}}^2-r_{\mathrm{odd}}^2-1 \big)(g_{Ls}^2+g_{Rs}^2) + 4 \sqrt{2} r_{\mathrm{odd}}\big(r_{\mathrm{even}}^2-r_{\mathrm{odd}}^2-1 \big)(g_{Ls} \ g_{Rs}), \\ \nonumber
x_3=& 0, \\ \nonumber
x_4=&-2 \sqrt{2}r_{\mathrm{odd}} (r_{\mathrm{odd}}(g_{Ls}^2+g_{Rs}^2)+2 g_{Ls}g_{Rs}),
\end{align}
where $g_{Ls}=g_{Rs}=-y \cos \theta /\sqrt{2}$ for the lightest $Z_2$-odd charged fermion and $g_{Ls}=g_{Rs}=y \sin \theta /\sqrt{2}$ for the heaviest $Z_2$-odd charged fermion.

\textbf{Vector Bosons}
\begin{align}
x_1=& 2\sqrt{2}((r_{\mathrm{even}}^4+4 \ r_{\mathrm{odd}}^2-r_{\mathrm{even}}^2(1+r_{\mathrm{odd}}^2))(g_{Lw}^2+g_{Rw}^2) ,\\ \nonumber
&-8 \ r_{\mathrm{odd}}(1-r_{\mathrm{even}}^2+r_{\mathrm{odd}}^2) g_{Lw} \ g_{Rw}, \\ \nonumber
x_2=& -2 \sqrt{2} (r_{\mathrm{odd}}^2 (-3-r_{\mathrm{even}}^2+r_{\mathrm{odd}}^2)(g_{Lw}^2+g_{Rw}^2) +8 \ r_{\mathrm{odd}}  \ g_{Lw} \ g_{Rw}) ,\\ \nonumber
x_3=& 8 \ \sqrt{2} (-1+r_{\mathrm{even}}^2)(g_{Lw}^2+g_{Rw}^2),
\\ \nonumber
x_4=&4 \ \sqrt{2} \ r_{\mathrm{odd}}(r_{\mathrm{odd}}(g_{Lw}^2+g_{Rw}^2)-4 \ g_{Lw} \ g_{Rw}),
\end{align}
where $g_{Lw}=g_{Rw}=-g_L \sin \theta /2$ for the lightest $Z_2$-odd charged fermion and $g_{Lw}=g_{Rw}=-g_L \cos \theta /2$ for the heaviest $Z_2$-odd charged fermion.

\bibliographystyle{apsrev4-1long}
\bibliography{darkmatter}

\end{document}